\documentclass[12 pt]{article}   
\usepackage{psfig,cite}       

\newenvironment{comment}[1]{}{}

\newcommand{\equ}[1]{(\ref{#1})}
\newcommand{\eqs}[1]{(\ref{#1})}

\newcommand{\ew}{electroweak~}
\newcommand{\non}{\nonumber}
\newcommand{\gsim}{\;\rlap{\lower 3.5 pt \hbox{$\mathchar \sim$}} \raise 1pt
 \hbox {$>$}\;}
\newcommand{\lsim}{\;\rlap{\lower 3.5 pt \hbox{$\mathchar \sim$}} \raise 1pt
 \hbox {$<$}\;}
\newcommand{\smallz}{{\scriptscriptstyle Z}} 
\newcommand{\smallw}{{\scriptscriptstyle W}} %

\newcommand{\smallS}{{\scriptscriptstyle S}}
\newcommand{\smallL}{{\scriptscriptstyle L}}
\newcommand{\smallR}{{\scriptscriptstyle R}}
\newcommand{\mz}{M_\smallz}
\newcommand{\mw}{M_\smallw}

\newcommand{\mt}{m_t}
\newcommand{\msbar}{\overline{\rm MS}}

\def\epe{\varepsilon'/\varepsilon}
\def\cO{{\cal O}}
\def\Heff{{\cal H}_{\rm eff}}
\def\f{\frac}
\def\cw{c_\smallw}
\def\cL{{\cal{L}}}
\def\cZ{{\cal{Z}}}\def\cB{{\cal{B}}}
\def\cW{{\cal{W}}}\def\cV{{\cal{V}}}
\def\cA{{\cal{A}}} \def\cS{{\cal{S}}}
\def\sws{\sin^2\theta_\smallw}
\def\swq{\sin^4\theta_\smallw}
\def\xmu{x_\mu}

\newcommand{\efe}[1]{Ref.\cite{#1}}

  \newenvironment{appendletterA}
 {
  \typeout{ Starting Appendix \thesection }
  \setcounter{equation}{0}
  
}{
  \typeout{Appendix done}
 }

\def\pl#1#2#3{{\it Phys. Lett. }{\bf B#1~}(19#2)~#3}

\def\pr#1#2#3{{\em Phys. Rev. }{\bf D#1~}(19#2)~#3}
\def\np#1#2#3{{\em Nucl. Phys. }{\bf B#1~}(19#2)~#3}

\newcommand{\be}{\begin{equation}}
\newcommand{\ee}{\end{equation}}
\newcommand{\een}{\end{subequations}}
\newcommand{\ben}{\begin{subequations}}
\newcommand{\beq}{\begin{eqalignno}}
\newcommand{\eeq}{\end{eqalignno}}
\newcommand{\bea}{\begin{eqnarray}}
\newcommand{\eea}{\end{eqnarray}}
\newcommand{\aem}{\alpha}

\def\as{\alpha_s}

\def \gev  {\mbox{ GeV}}

\def \ms   {\overline{\mbox{MS}}}

\def\epe{\varepsilon'/\varepsilon}

\newcommand{\mb}{m_{\rm b}}
\newcommand{\ord}{{\cal O}}
\newcommand{\aw}{\alpha_\smallw}

\begin{document}              

\begin{titlepage}
\begin{flushright}
        \small
        TUM-HEP-361/99\\
        MPI-PhT/99-53\\
        October 1999
\end{flushright}

\begin{center}
\vspace{2cm}
{\Large\bf Electroweak Penguin Contributions to\\
\vspace{1mm}
Non-Leptonic $\Delta F = 1$ 
Decays at NNLO} 

\vspace{0.5cm}
\renewcommand{\thefootnote}{\fnsymbol{footnote}}
{\bf          Andrzej J.~Buras$^1$, Paolo
	      Gambino$^{1}\footnote[2]{Address after November 1,
	      1999, Theory Division, CERN, CH-1211 Geneva 23,
              Switzerland.}$ and Ulrich A. Haisch$^{1, 2}$
}
\setcounter{footnote}{0}
\vspace{.8cm}

{\it $^1$Technische Universit{\"a}t M{\"u}nchen,\\
                       Physik Dept., D-85748 Garching, Germany }
\vspace{2mm}

{\it $^2$Max Planck Institut f{\"u}r Physik -- Werner-Heisenberg-Institut,\\
        F{\"o}hringer Ring 6, D-80805 Munich,  Germany}
\vspace{3.cm}

{\large\bf Abstract}
\end{center}
We calculate the $\ord(\alpha_s)$ corrections to 
the $Z^0$-penguin and electroweak box diagrams relevant for 
non-leptonic $\Delta F=1$ decays with $F=S,B$. 
This calculation provides the complete $\ord (\aw \as)$ 
and $\ord (\aw \as \sin^2\theta_\smallw\mt^2)$ corrections  
($\aw = \aem/\sws$)
to the Wilson coefficients of the electroweak penguin four
quark operators relevant for non-leptonic K- and B-decays. We argue
that this is the dominant part of the next-next-to-leading (NNLO)
contributions to these 
coefficients. Our results allow to reduce considerably the
uncertainty due to the definition  of the top
quark mass present in the existing NLO calculations of non-leptonic 
decays. The NNLO corrections to the coefficient of the color singlet 
$(V-A)\otimes (V-A)$ \ew penguin operator $Q_9$ relevant for $B$-decays
are generally moderate, amount to a few percent for the choice 
$\mt (\mu_t = \mt)$ and depend only weakly on the renormalization scheme.
Larger NNLO corrections with substantial scheme dependence 
are found for the coefficients of the remaining \ew penguin operators $Q_7$,
$Q_8$ and $Q_{10}$. In particular, the strong scheme dependence of the NNLO 
corrections to $C_8$ allows to reduce considerably the scheme dependence
of $C_8 \langle Q_8 \rangle_2$ relevant for the ratio $\varepsilon'/\varepsilon$.

\noindent


\end{titlepage}
\newpage

\section{Introduction}
Electroweak penguin operators govern rare semi-leptonic decays
such as $K\to\pi\nu\bar\nu$, $K_L\to\pi^0e^+e^-$, $B\to\pi\nu\bar\nu$
and $B_{s,d}\to\mu\bar\mu$ and contribute sometimes in an important
manner to non-leptonic K- and B- decays. Among the latter one should
mention the CP-violating ratio $\epe$ in $K_L\to\pi\pi$ decays
and electroweak contributions to two-body decays like $B\to\pi K$,
$B_s\to\pi^0\phi$ etc. \cite{fleischer}.

The effective weak Hamiltonian for these decays has the following generic
structure \cite{buchalla:96}
\begin{equation}\label{b1}
{\cal H}_{eff}=\frac{G_F}{\sqrt{2}}\sum_i V^i_{\rm CKM}C_i(\mu)Q_i~.
\end{equation}
Here $G_F$ is the Fermi constant and $Q_i$ are the relevant local
operators which, in addition to electroweak penguin operators,
include current-current operators, QCD-penguin operators and magnetic
penguin operators. The Cabibbo-Kobayashi-Maskawa
factors $V^i_{CKM}$ and the Wilson coefficients $C_i$ describe the 
strength with which a given operator enters the Hamiltonian.
The decay amplitude for a decay of a meson M to a final state F
is simply given by $\langle F|{\cal H}_{eff}|M\rangle$.

The renormalization scale ($\mu$) dependence of $\vec C^T(\mu)=
(C_1(\mu),...)$ is governed by the renormalization group equation (RGE)
whose solution is given by
\begin{equation}\label{b2}
\vec C(\mu)=
\left[
T_g \exp\int^{g_s(\mu)}_{g_s(\mu_W)}dg_s'
{\hat\gamma^T(g_s',\alpha)\over\beta(g_s')}
\right] \vec C(\mu_W) \, , 
\end{equation}
where $T_g$ is the $g$-ordering operator, $\mu_W=\ord(\mw)$
and $\mu$ is $\ord(\mb)$ and $\ord(1 \gev)$ for B-decays and K-decays
respectively. $\beta(g_s)$ governs the evolution of the QCD coupling
constant $g_s$ and $\hat\gamma$ is the anomalous dimension matrix
which depends on the QED coupling constant $\alpha$ in addition
to $\alpha_s=g_s^2/4\pi$. In what follows we will work to first
order in $\alpha$.

Now, the initial conditions $\vec C(\mu_W)$ are linear combinations of
the so called Inami-Lim functions \cite{inami:81} such as $B_0(x_t)$ with
$x_t=\mt^2/\mw^2$ resulting from
box diagrams, $C_0(x_t)$ from $Z^0$-penguin diagrams,
$D_0(x_t)$ from the photon penguin diagrams, $E_0(x_t)$                   
from the gluon penguin diagrams etc. The full list of functions
including also those relevant for $\Delta S=2$, $\Delta B=2$
transitions and for radiative  B- decays can be found in
\cite{AJBLH}. We will give explicit expressions for some of these
functions below.

As shown in \cite{PBE0} any decay amplitude can then be
written as a linear combination of the Inami-Lim functions
to be denoted by $F^{(0)}_r(x_t)$
\begin{equation}
A({\rm decay}) = P_0({\rm decay}) + 
\sum_r P_r({\rm decay}) \, F^{(0)}_r(x_t) \, ,
\label{b3}
\end{equation}
where the sum runs over all functions contributing to a given decay
and $P_0$ summarizes contributions from internal up and charm quarks. 
The process dependent 
coefficients $P_r$ include the effect of the renormalization
group evolution from $\mu_W$ down to $\mu$ given in (\ref{b2})
as well the matrix elements of the operators $Q_i$. On the other
hand the Inami-Lim functions $F^{(0)}_r(x_t)$ are process
independent. That is for instance the functions $B_0$ and $C_0$
enter both the semi-leptonic rare decays $K\to\pi\nu\bar\nu$ and $\epe$
which is related to non-leptonic decays $K_L\to\pi\pi$.

This {\it Penguin-Box Expansion} 
is very well suited for the study of the extensions of the
Standard Model (SM) in which new particles are exchanged in the loops.
We know already that these particles are relatively heavy 
and consequently they can be integrated out together with
the weak bosons and the top quark. If there are no new local operators
the mere change is to modify the functions $F^{(0)}_r(x_t)$ which now
acquire the dependence on the masses of new particles such as
charged Higgs bosons and supersymmetric partners. The process
dependent coefficients $P_0$ and $P_r$ remain unchanged unless
new effective operators with different Dirac and color structures
have to be introduced.

Now, the universal Inami-Lim functions $F^{(0)}_r(x_t)$ result
from one-loop box and penguin diagrams without QCD corrections
and it is of interest to ask how these functions are modified
when $\ord(\alpha_s)$ corrections to box and penguin diagrams
are included. 

The interest in answering this question is as follows:
\begin{itemize}
\item
The estimate of the size of QCD corrections to
 the relevant decay branching ratios.
\item
The reduction of various unphysical scale dependences. In the
case studied in this paper, this is in particular the
dependence of QCD corrections on the scale $\mu_t$ at which
the running top quark mass is defined. As some of the functions
$F^{(0)}_r(x_t)$ depend strongly on $\mt$, their $\mu_t$
dependence may result in the uncertainties as large
as $\pm 15\%$ in the corresponding branching ratios.
Only by calculating QCD corrections to box and penguin
diagrams can this dependence be reduced.
\item
The universality of the top dependent functions 
can be violated by $\ord(\alpha_s)$ corrections.
For instance in the case of
semi-leptonic FCNC transitions there is no gluon exchange in
a $Z^0$-penguin diagram parallel to the $Z^0$-propagator but
such an exchange takes place in non-leptonic decays in which all
external particles are quarks. 
The same applies to box diagrams contributing to $K\to\pi\nu\bar\nu$
and $\epe$ respectively. It is of interest then to find out
whether the breakdown of the universality is substantial.
\item Most importantly, however, the inclusion of $\ord(\as)$ corrections
to penguin and box diagrams relevant for non-leptonic decays justifies
the simultaneous inclusion of particular next-next-to-leading (NNLO) QCD
corrections to the renormalization group transformation in (\ref{b2}). In the
case of the \ew penguin operator $Q_8$ (see (\ref{OS4})), relevant for 
$\varepsilon'/\varepsilon$, this results in a welcome renormalization scheme
dependence of $C_8(\mu)$ which in turn allows to reduce considerably the
renormalization scheme dependence of $\varepsilon'/\varepsilon$
present at NLO.

\end{itemize}

So far the following $\ord(\alpha_s)$ corrections to box and penguin
diagrams have been calculated:

\begin{itemize}
\item
$\ord(\alpha_s)$ corrections to $Z^0$-penguin function $C_0$ and
to the box-diagram function $B_0$ in the case of rare semi-leptonic
decays like $K\to \pi\nu\bar\nu$, $B\to\mu\bar\mu$ etc. 
\cite{buchalla:93a,buchalla:93b,misiak99}
\item
$\ord(\alpha_s)$ corrections to chromomagnetic and magnetic penguins
relevant for radiative decays $B\to X_s\gamma$ and $B\to X_s l^+l^-$
\cite{Yao1,GH97,BKP2,GAMB}.
\item
$\ord(\alpha_s)$ corrections to the photon penguin function $D_0$
relevant for $B\to X_s l^+l^-$ and $K_L\to\pi^0 l^+l^-$ \cite{misiak}.
\item 
$\ord(\alpha_s^2)$ corrections to the matching conditions of all the 
 operators $Q_{1-6}$, including the $\ord(\alpha_s)$ contributions
to the gluon penguin function $E_0$ \cite{misiak}.

\end{itemize}

The purpose of the present paper is the calculation of
$\ord(\alpha_s)$ corrections to $Z^0$-penguin and box-diagrams 
relevant for non-leptonic decays, such as two-body B-meson decays
and $K_L\to\pi\pi~(\epe)$. This will allow to reduce the
$\mu_t$-dependence in the NLO expressions present in the
literature, to investigate the breakdown of the universality
of the relevant Inami-Lim functions and to study the renormalization
scheme dependence at the NNLO level.

As we will discuss explicitly in Section 3 the QCD corrections
calculated here are a part of the complete next-next-to-leading
(NNLO) corrections to non-leptonic decays in a renormalization
group improved perturbation theory. In order to complete
the NNLO calculations of the relevant Wilson coefficients
one would have to calculate $\ord(\alpha_s^2)$ corrections to QCD penguin
diagrams and in particular perform three-loop calculations
$\ord(\alpha_s^3)$, $\ord(\alpha\alpha_s^2)$ of the
anomalous dimensions of the full set of  operators,
which is a formidable task and clearly beyond the 
scope of our paper. However our calculation is sufficient to obtain
the complete $\ord (\aw \as)$ and the $\ord (\aw \as \sws \mt^2)$
corrections to the Wilson coefficients $C_{7-10} (\mu)$ of
the electroweak penguin operators where $\aw = \aem/\sws$. It is
also sufficient to investigate the issue of the $\mu_t$-dependence of
these coefficients and of its reduction through $\ord (\as)$
corrections calculated here. Finally it allows to analyze the breakdown
of the universality in the Inami-Lim functions related to
$Z^0$-penguin and box diagrams.  

Our paper is organized as follows. In Section 2 we recall the
known effective Hamiltonian for $\Delta S=1$ decays at the
NLO level. We list the contributing operators and give the
expressions for the Inami-Lim functions. In Section 3 we
discuss  our paper in the context of a
complete NNLO calculation, we motivate our approximations and
we outline the strategy. In Section 4 we elaborate on the renormalization
scheme dependence. The calculation of the gluonic
corrections to the $Z^0$ penguin diagrams  and to the \ew box diagrams
is described in Sections \,\ref{Zpeng} and \ref{boxes}, respectively. In
Section 7 we collect the results in terms of $\ord(\alpha_\smallw\as)$ 
contributions to the Wilson coefficients $C_{7-10}(\mw)$ and discuss 
their numerical relevance at various scales, as well as the residual 
scale and scheme dependences. Finally, in Section 8 we summarize our paper
and we briefly discuss the impact of our findings on the phenomenology of
non-leptonic decays.

\section{Notation and Conventions}
\setcounter{equation}{0}
In this section we establish our notation and recall some definitions that
will be useful in the rest of the paper. We give explicit formulae for $\Delta
S=1$ decays. It is straightforward to transform them to the $\Delta B=1$ case.
The effective Hamiltonian for $\Delta S=1$
transitions can be written as \cite{buchalla:96}:
\begin{equation}
\Heff(\Delta S=1) = 
-\frac{G_{\rm F}}{\sqrt{2}}\, V_{ts}^* V_{td}^{} \,\sum_{i=1}^{10}
C_i(\mu)\,Q_i(\mu), 
\label{eq:HeffdF1:1010}
\end{equation}
where we have dropped the terms proportional to $ V_{us}^* V_{ud}$
which are of no concern to us here. In \cite{buchalla:96} $C_i(\mu)\equiv 
y_i(\mu)$.
The operators $Q_i$  are given explicitly  as follows:

{\bf Current--Current :}
\begin{equation}\label{OS1} 
Q_1 = (\bar s_{\alpha} u_{\beta})_{V-A}\;(\bar u_{\beta} d_{\alpha})_{V-A}
~~~~~~Q_2 = (\bar s u)_{V-A}\;(\bar u d)_{V-A} 
\end{equation}

{\bf QCD--Penguins :}
\begin{equation}\label{OS2}
Q_3 = (\bar s d)_{V-A}\sum_{q=u,d,s}(\bar qq)_{V-A}~~~~~~   
 Q_4 = (\bar s_{\alpha} d_{\beta})_{V-A}\sum_{q=u,d,s}(\bar q_{\beta} 
       q_{\alpha})_{V-A} 
\end{equation}
\begin{equation}\label{OS3}
 Q_5 = (\bar s d)_{V-A} \sum_{q=u,d,s}(\bar qq)_{V+A}~~~~~  
 Q_6 = (\bar s_{\alpha} d_{\beta})_{V-A}\sum_{q=u,d,s}
       (\bar q_{\beta} q_{\alpha})_{V+A} 
\end{equation}

{\bf Electroweak--Penguins :}
\begin{equation}\label{OS4} 
Q_7 = {3\over 2}\;(\bar s d)_{V-A}\sum_{q=u,d,s}e_q\;(\bar qq)_{V+A} 
~~~~~ Q_8 = {3\over2}\;(\bar s_{\alpha} d_{\beta})_{V-A}\sum_{q=u,d,s}e_q
        (\bar q_{\beta} q_{\alpha})_{V+A}
\end{equation}
\begin{equation}\label{OS5} 
 Q_9 = {3\over 2}\;(\bar s d)_{V-A}\sum_{q=u,d,s}e_q(\bar q q)_{V-A}
~~~~~Q_{10} ={3\over 2}\;
(\bar s_{\alpha} d_{\beta})_{V-A}\sum_{q=u,d,s}e_q\;
       (\bar q_{\beta}q_{\alpha})_{V-A} \,.
\end{equation}
Here, $e_q$ denotes the electrical quark charges reflecting the
electroweak origin of $Q_7,\ldots,Q_{10}$. 
The initial conditions for  the Wilson coefficients $C_i$ 
at $\mu=\mw$ obtained from the one-loop matching  of the full to the effective
theory are given in the NDR renormalization scheme as follows \cite{BJL93}: 
\begin{eqnarray}
C_1(\mw) &=&     \frac{11}{2} \; \frac{\as(\mw)}{4\pi} \, ,
\label{eq:CMw1} \\
C_2(\mw) &=& 1 - \frac{11}{6} \; \frac{\as(\mw)}{4\pi}
               - \frac{35}{18} \; \frac{\aem}{4\pi} \, ,
\label{eq:CMw2} \\
C_3(\mw) &=& -\frac{\as(\mw)}{24\pi} \widetilde{E}_0(x_t)
             +\frac{\alpha_\smallw}{6\pi} 
             \left[ 2 B_0(x_t) + C_0(x_t) \right] \, , 
\label{eq:CMw3} \\
C_4(\mw) &=& \frac{\as(\mw)}{8\pi} \widetilde{E}_0(x_t) \, ,
\label{eq:CMw4} \\
C_5(\mw) &=& -\frac{\as(\mw)}{24\pi} \widetilde{E}_0(x_t) \, ,
\label{eq:CMw5} \\
C_6(\mw) &=& \frac{\as(\mw)}{8\pi} \widetilde{E}_0(x_t) \, ,
\label{eq:CMw6} \\
C_7(\mw) &=& \frac{\alpha_\smallw}{6\pi} \sws\left[ 4 C_0(x_t) + \widetilde{D}_0(x_t)
\right]\, ,
\label{eq:CMw7} \\
C_8(\mw) &=& 0 \, ,
\label{eq:CMw8} \\
C_9(\mw) &=& \frac{\alpha_\smallw}{6\pi} 
\left[ \sws (4 C_0(x_t) + \widetilde{D}_0(x_t)) +
              10 B_0(x_t) - 4 C_0(x_t) \right] \, ,
\label{eq:CMw9} \\
C_{10}(\mw) &=& 0 \, ,
\label{eq:CMw10} 
\end{eqnarray}
where we have introduced $\alpha_\smallw= g^2/4\pi= \aem/\sws$. $g$ is the 
weak coupling of $SU(2)_L$.
We recall that 
\begin{eqnarray}
B_0(x_t) &=& \frac{1}{4} \left[ \frac{x_t}{1-x_t} 
+ \frac{x_t \ln x_t}{(x_t-1)^2}
\right]\, , \label{eq:B0} \\
C_0(x_t) &=& \frac{x_t}{8} \left[\frac{x_t-6}{x_t-1} 
+ \frac{3 x_t + 2}{(x_t-1)^2}
\ln x_t \right]\, ,
\label{eq:C0} \\
D_0(x_t) &=& -\frac{4}{9} \ln x_t + 
\frac{-19 x_t^3 + 25 x_t^2}{36 (x_t-1)^3} +
\frac{x_t^2 (5 x_t^2 - 2 x_t - 6)}{18 (x_t-1)^4} \ln x_t \, ,
\label{eq:Dxt} \\
E_0(x_t) &=& -\frac23 \ln x_t + \frac{x_t (18-11x_t -x_t^2)}{12 (1-x_t)^3}
+ \frac{x_t^2(15-16x_t +4 x_t^2)}{6(1-x_t)^4}\, ,
\\
\widetilde{D}_0(x_t) &=& D_0(x_t) - \frac{4}{9}\, , \ \ \ \ \ \ \ \ \ 
\widetilde{E}_0(x_t) = E_0(x_t) - \frac{2}{3} 
\, .
\label{eq:Dxttilde} 
\end{eqnarray}
$B_0(x_t)$
results from the evaluation of the box diagrams, $C_0(x_t)$ from the
$Z^0$-penguin,  $D_0(x_t)$ from the photon penguin diagrams
and $E_0(x_t)$ from QCD penguin diagrams.
The constants $-4/9$ and $-2/3$ in (\ref{eq:Dxttilde})
are characteristic for the NDR scheme. They are absent in the HV scheme.
For $\mu\not=\mw$ non-vanishing
$C_8$ and $C_{10}$ are generated through QCD effects.
The formulae (\ref{eq:CMw1})-(\ref{eq:CMw10}) apply also to the $\Delta B=1$
case with the appropriate change of fields in $Q_i$.

Let us next recall that in the leading order (LO) of the
renormalization group improved perturbation theory in which $(\as t)^n$
and $\aem t (\as t)^n$ terms with $t = \ln \, (\mw^2/\mu^2)$ are summed
only $C_2 (\mw) = 1$ is different from zero. In particular $C_{7-10}
(\mw) = 0$. The initial conditions given in
(\ref{eq:CMw1})-(\ref{eq:CMw10}) are appropriate to next-to-leading
order (NLO) in which $\as (\as t)^n$ and $\aem (\as t)^n$ terms are
summed. In the next-next-to-leading order (NNLO) in which $\as^2 (\as
t)^n$ and $\aem \as (\as t)^n$ are summed, $\ord (\as^2)$ terms in the
initial conditions of $C_{1-6} (\mw)$ and $\ord (\aem \as)$ terms for
all $C_i (\mw)$ have to be included. In the present paper we will
calculate the dominant $\ord (\aem \as)$ corrections to the
coefficients $C_{7-10} (\mw)$ of the electroweak penguin operators. As
we will discuss below this will be sufficient to sum the dominant
contribution of the $\aem \as (\as t)^n$ logarithms.

Finally we should stress that among $\ord (\aem \as)$ terms we
distinguish between $\ord (\aw \as)$ and $\ord (\aw \as \sws)$ terms
for reasons to be explained in detail below. 

\section{General Structure at NNLO and Strategy}
\label{strategy}
\setcounter{equation}{0}
Our aim is to compute the $\ord(\as)$ corrections to the $Z^0$-penguin
diagrams and electroweak box diagrams  relevant for non-leptonic
$\Delta F=1$ decays. This calculation constitutes only a part of the
complete computation 
of the Wilson coefficients $C_i(\mu)$ (i=1,...10) at NNLO in the
renormalization group improved perturbation theory. On the other
hand, as we will now demonstrate, our results combined with the
known $\ord(\as)$ and $\ord(\as^2)$ anomalous dimensions of $Q_i$
provide the complete $\ord(\aw\as)$ corrections to the Wilson coefficients
$C_{7-10}(\mu)$ of the electroweak penguin operators not suppressed by
$\sws$ and the $\ord (\aw \as \sws)$ corrections quadratic in $\mt$ to
these coefficients. These corrections turn out to be by far the
dominant contributions to $C_{7-10} (\mu)$ at the NNLO level.

In order to prove these statements it is instructive to describe the
computation of the Wilson coefficients $C_i(\mu)$ including LO, NLO
and NNLO corrections. Generalizing the standard procedure at NLO
\cite{BJL93,AJBLH} to include NNLO corrections we proceed as follows.

\vspace{2mm}
{\bf Step 1:} An amplitude for a properly chosen non-leptonic quark
decay is calculated perturbatively in the full theory including all
sorts of diagrams such as QCD penguin diagrams, electroweak penguin
diagrams, box diagrams, W-boson exchanges and QCD corrections
to all these diagrams. The result including LO, NLO and 
NNLO correction
is given schematically as follows:
\begin{eqnarray}\label{full}
A_{full}
&\!\!=\!\!&
\langle\vec Q^{(0)}\rangle^T
\left[\vec A^{(0)}+\frac{\as(\mw)}{4\pi}\vec A_s^{(1)} +
\left(\frac{\as(\mw)}{4\pi}\right)^2
\vec A_s^{(2)} 
+\frac{\aem}{4\pi}\vec A_e^{(1)}+
\frac{\aem}{4\pi}\frac{\as(\mw)}{4\pi}\vec A_{es}^{(2)}\right]
\nonumber \\
&\!\!\equiv\!\!& 
\langle\vec Q(\mw)\rangle^T \vec C(\mw) \, , 
\end{eqnarray}
where $\langle \vec Q^0\rangle$ is a ten dimensional column vector
built out of tree level matrix elements of the  operators $Q_i$.
The superscripts (0), (1) and (2) denote LO, NLO and NNLO contributions,
respectively. The $\ord (\aem \as)$ corrections include $\ord
(\aw \as)$, $\ord (\aw \as \sws)$ and $\ord (\aw \as \swq)$ terms. 

\vspace{2mm}
{\bf Step 2:} In order to extract the coefficients $\vec C(\mw)$ from
(\ref{full}) one has to calculate the matrix elements of $Q_i$ between
the same external quark states as in Step 1. This involves generally the
computation of the operator insertions into the current-current,
gluon penguin and photon penguin diagrams of the effective theory
(W, top and $Z^0$ have been integrated out) together with QCD and QED
corrections to these insertions. Including LO, NLO and NNLO corrections 
one finds 
\begin{eqnarray}\label{effective}
\langle\vec Q (\mw) \rangle
&\!\!=\!\!&
\left[ \hat 1+\frac{\as(\mw)}{4\pi}\hat r_s^{(1)} +
\left(\frac{\as(\mw)}{4\pi}\right)^2
\hat r_s^{(2)} 
+\frac{\aem}{4\pi}\hat r_e^{(1)}+
\frac{\aem}{4\pi}\frac{\as(\mw)}{4\pi}\hat r_{es}^{(2)}\right]
\langle\vec Q^{(0)}\rangle
\nonumber \\
\end{eqnarray}
with $\hat r_i$  being $10\times 10$ matrices. As $W$ and $Z^0$ have been
integrated out only $\ord (\aw \as \sws)$ terms are present in $\ord
(\aem \as)$ corrections. 

\vspace{2mm}
{\bf Step 3:} From (\ref{full}) and (\ref{effective}) we extract

\begin{eqnarray}\label{cw}
\vec C (\mw)
&=&
\vec C^{(0)}+\frac{\as(\mw)}{4\pi}\vec C_s^{(1)} +
\left(\frac{\as(\mw)}{4\pi}\right)^2
\vec C_s^{(2)}
+\frac{\aem}{4\pi}\vec C_e^{(1)}+
\frac{\aem}{4\pi}\frac{\as(\mw)}{4\pi}\vec C_{es}^{(2)} \, , \hspace{0.75cm}
\end{eqnarray}
where
\begin{eqnarray}\label{cwi}
\vec C^{(0)} &=& \vec A^{(0)} \hspace{2mm} = \hspace{2mm} (0, 1, 0,
\ldots, 0)^T \, , \\ 
\vec C_s^{(1)}&=& \vec A_s^{(1)}-\hat r_s^{(1)^T} \vec C^{(0)} \, , \\
\vec C_e^{(1)}&=& \vec A_e^{(1)}-\hat r_e^{(1)^T} \vec C^{(0)} \, , \\
\vec C_s^{(2)}&=& \vec A_s^{(2)}-\hat r_s^{(2)^T} \vec C^{(0)}  
-\hat r_s^{(1)^T} \vec C_s^{(1)} \, , 
\\
\vec C_{es}^{(2)}&=& \vec A_{es}^{(2)}-\hat r_s^{(1)^T} \vec C_e^{(1)}
-\hat r_{es}^{(2)^T} \vec C^{(0)}
-\hat r_e^{(1)^T} \vec C_s^{(1)} \, .
\label{ces}
\end{eqnarray}

The electroweak penguin components of $\vec C_e^{(1)}$ and the
components of $\vec C_s^{(1)}$ which contribute to the coefficients of
the operators $Q_{1-6}$ are given explicitly in
(\ref{eq:CMw1})-(\ref{eq:CMw10}). $\vec C_s^{(2)}$ has been calculated
in \cite{misiak}, but it
contributes only to $\ord (\as^2)$ corrections to the coefficients of
$Q_{1-6}$. 

Among the four terms contributing to $\vec C_{es}^{(2)}$ only $\vec
A_{es}^{(2)}$ and $\hat r_s^{(1)^T} \vec C_e^{(1)}$ are of interest to
us as these are the only ones contributing to $\ord (\aw \as)$
corrections and to $\ord (\aw \as \sws)$ corrections quadratic in
$\mt$ which we aim to calculate. The third and fourth term in
(\ref{ces}) contribute only to $\ord (\aw \as \sws)$ corrections. The
third term is $\mt$-independent and  unknown. The last term can be
extracted from the known one-loop results but it has no dependence 
quadratic in $\mt$ and will not be included here. 

The main purpose of this paper is then the two-loop calculation of
$Z^0$-penguin and box diagrams giving $\vec A_{es}^{(2)}$. The
contribution from the effective theory $\hat r_s^{(1)^T} \vec
C_e^{(1)}$ can be extracted from the known one-loop
results. In the case of the Wilson coefficients of electroweak penguin
operators a simplification occurs as only the operator insertions in
current-current topologies in $\hat r_s^{(1)^T}$ contribute. Were we
interested also in the coefficients of the QCD-penguin operators, also
the insertions into QCD-penguin and QED-penguin topologies would have
to be retained.

\vspace{2mm}
{\bf Step 4:} We use next the renormalization group transformation to
find 

\begin{equation}\label{cmu}
\vec C (\mu) = \hat U(\mu,\mw,\alpha)\vec C(\mw) \, , 
\end{equation}
where

\begin{equation}\label{utot}
\hat U(\mu,\mw,\alpha)=\hat U(\mu, \mw)+ \frac{\aem}{4\pi}
\left[\hat R^{(0)} (\mu, \mw) +\hat R^{(1)} (\mu, \mw) +\hat R^{(2)}
(\mu, \mw) \right]
\end{equation}
with the pure QCD evolution given by
\begin{equation}\label{uQCD}
\hat U(\mu,\mw)=\hat U^{(0)}(\mu, \mw)+
\hat U^{(1)}(\mu, \mw)+\hat U^{(2)}(\mu, \mw).
\end{equation}
The matrices $\hat U^{(i)}$ and $\hat R^{(i)}$ are 
functions of the anomalous dimension matrices of
the operators in question and of the QCD beta function.
Explicit expressions for $\hat U^{(0)}$, $\hat U^{(1)}$,
$\hat R^{(0)}$ and $\hat R^{(1)}$ can be extracted from
\cite{BJL93,ciuchini:94}. $\hat R^{(2)}$ and  $\hat U^{(2)}$ are not known
as they require the evaluation of the three-loop anomalous
dimension matrices $\ord(\aem\as^2)$ and $\ord(\as^3)$,
respectively. From the point of view of the expansion in $\as$
in the renormalization group improved perturbation theory,
$\hat U^{(0)}$, $\hat U^{(1)}$ and $\hat U^{(2)}$ are
$\ord(1)$, $\ord(\as)$ and $\ord(\as^2)$, respectively.
$\hat R^{(0)}$, $\hat R^{(1)}$  and $\hat R^{(2)}$ 
are $\ord(1/\as)$, $\ord(1)$ and $\ord(\as)$, respectively.

Inserting (\ref{utot}) and (\ref{cw}) into (\ref{cmu})
and expanding in $\as$ we find

\begin{equation}\label{cfin}
\vec C (\mu) = \vec C_s(\mu)+ 
\frac{\alpha}{4\pi}\left[\vec C_I(\mu)+\vec C_{II}(\mu)\right]~.
\end{equation}
$\vec C_s(\mu)$ results from $\ord(1)$, $\ord(\as)$ and $\ord(\as^2)$
terms in $\vec C (\mw)$ and the QCD evolution $\hat U(\mu,\mw)$.
$\vec C_I(\mu)$ results from terms $\ord(\aem)$ and $\ord(\aem\as)$ 
in $\vec C(\mw)$ and $\hat U(\mu,\mw)$. Finally $\vec C_{II}(\mu)$
is found by taking the contributions
$\ord(1)$, $\ord(\as)$ and $\ord(\as^2)$ in $\vec C(\mw)$ and
performing renormalization group transformation using $\hat R^{(i)}$.
Explicitly we have:

\begin{eqnarray}\label{cs}
\vec C_s(\mu)
&=&
\hat U^{(0)}(\mu,\mw) \left[\vec C^{(0)}+\frac{\as(\mw)}{4\pi}
\vec C_s^{(1)} +\left(\frac{\as(\mw)}{4\pi}\right)^2
\vec C_s^{(2)}\right] 
\nonumber\\
&+&
\hat U^{(1)}(\mu,\mw) \left[\vec C^{(0)}+\frac{\as(\mw)}{4\pi}
\vec C_s^{(1)}\right] + \hat U^{(2)}(\mu,\mw) \vec C^{(0)} \, , 
\end{eqnarray}

\begin{equation}\label{cI}
\vec C_I(\mu)=
\hat U^{(0)}(\mu,\mw) \left[\vec C_e^{(1)}+\frac{\as(\mw)}{4\pi}
\vec C_{es}^{(2)}\right]+\hat U^{(1)}(\mu,\mw) \vec C_e^{(1)} \, , 
\end{equation}

\begin{eqnarray}\label{cII}
\vec C_{II}(\mu)
&=&
\hat R^{(0)}(\mu,\mw) \left[\vec C^{(0)}+\frac{\as(\mw)}{4\pi}
\vec C_s^{(1)} +\left(\frac{\as(\mw)}{4\pi}\right)^2
\vec C_s^{(2)}\right]
\nonumber\\
&+&
\hat R^{(1)}(\mu,\mw) \left[\vec C^{(0)}+\frac{\as(\mw)}{4\pi}
\vec C_s^{(1)}\right] + \hat R^{(2)}(\mu,\mw) \vec C^{(0)} \, . 
\end{eqnarray}

Let us now identify the $\ord(\aw\as)$ contributions to electroweak
penguin coefficients, calculated in subsequent sections, in this full
NNLO result. They are fully contained in the last two terms in
$\vec C_{I}(\mu)$. These two terms can be schematically decomposed as
follows:
\begin{equation}
F_1(x_t,\as)+ F_2(x_t,\as) \, \sws \, . 
\end{equation}
The $\ord(\aw\as)$ corrections are represented by the first term and
our calculation provides the complete result for $F_1(x_t,\as)$.
On the other hand, our calculation gives only a partial result for 
$F_2(x_t,\as)$ which is $\ord(\aw \as \sws)$. The contributions from
gluon corrections to photon penguin diagrams and $\ord (\aem)$
corrections to QCD penguin diagrams 
which both contribute to
$F_2 (x_t, \as)$ are still missing in the case of non-leptonic
decays. Similarly, some $\ord(\aw \as \sws)$ corrections contributing to the 
Wilson coefficient functions through $\vec C_{II}(\mu)$ are not 
known. Yet, as we will argue below all these contributions
to the Wilson coefficients of electroweak penguin operators are
expected to be much smaller than the $\ord(\aw\as)$ contributions 
calculated by us. Needless to say there are no contributions to
$C_{7-10}$ contained in $\vec C_s(\mu)$.

In order to understand better the dominance of $\ord (\aw \as)$ over 
$\ord (\aw \as \sws)$ corrections, let us look at the NLO result, where
the issue concerns the dominance of $\ord (\aw)$ corrections over 
$\ord (\aw \sws)$ corrections. From the expressions given in the previous 
section we see that at NLO  $C_9(\mw)$  is much larger than 
$ C_7(\mw)$: in units of ${\alpha_\smallw}/{6\pi}$ we have 
$C_9(\mw)=-4.46$ and   $C_7(\mw)=+0.52$.
More precisely,  $C_9(\mw)= 2.24 \, \sws - 4.97$,
namely at the electroweak scale $C_9$ is dominated by the second term, 
unsuppressed by $\sws$, 
while the first one accounts for  10\% of the total. 
The dominant term includes the box diagrams and the 
$SU(2)_L$ component of the $Z^0$ penguin diagrams, all contributing 
$\ord(g^2)$ terms unsuppressed by $\sws$: it 
can be called   the  {\it purely weak} contribution.

As we discussed above, the complete QCD corrections to the
coefficients of the  \ew penguin operators involve the computation of
the gluonic corrections to the one-loop $Z^0$ and photon penguins and 
to the \ew boxes. The two-loop $Z^0$ penguins and boxes can be calculated 
at $\ord(\alpha\as)$ setting all external momenta to zero. 
They   are entirely responsible for 
the {\it purely weak} contribution to $C_9(\mw)$, 
which is largely dominant at the one-loop level, as we have just seen.
On the other hand, the calculation of the two-loop
photon penguins is more involved, essentially  because the corresponding 
diagrams lack  a heavy mass scale like $\mz$ for the 
$Z^0$ penguins. Very recently, the color singlet component of this class
of diagrams has been computed in a different context \cite{misiak}:  we have
verified (see Section\,\ref{res}) that its $\ord(\aw \as \sws)$ 
contribution to $C_{7,9}$ is small compared to the one of the boxes 
and of the $Z^0$ penguin. 
The color octet component has not yet been calculated.
One possible  strategy  therefore consists in  
computing the gluonic corrections to $Z^0$ penguin and \ew box diagrams exactly
at  $\ord(\alpha_\smallw\as)$ and in neglecting all 
corrections vanishing as $\sws \rightarrow 0$, in particular 
corrections to the photon penguin. 
These {\it purely weak} contributions form a gauge-independent subset.

Before embarking in a complex two-loop calculation, it is also interesting to
see how a Heavy Top Expansion (HTE), i.e.\ an expansion in
inverse powers of the top quark mass, could approximate the complete result.
We notice that at the one-loop level 
the only contributions which are quadratic in the top 
quark mass originate from the $Z^0$ penguin diagrams, i.e.\ from $C_0$. 
This feature persists at the two-loop level $\ord(\alpha_\smallw\as)$:
restricting our analysis to these potentially 
enhanced contributions would simplify 
significantly our task. However, a closer look at the one-loop Wilson 
coefficients shows that, despite the fact that 
the HTE of $B_0, C_0$ and $\widetilde{D}_0$ 
converges rapidly, the leading order of the HTE approximates well 
$C_7(\mw)$ but not  $C_9(\mw)$ (it gives $-1.64 
\,\f{\alpha_\smallw}{6\pi}$ instead of $-4.45\,\f{\alpha_\smallw}{6\pi}$; 
this is due to the large coefficient in front of $B_0$ in (\ref{eq:B0}),
which has no quadratic term in $\mt$).
It is therefore unlikely that the leading order of the HTE  
provides by itself a good approximation at the two-loop level.
On the other hand, keeping all terms unsuppressed by $\sws$
(i.e.\ the {\it purely weak} ones) 
together with the leading HTE of the rest in the one-loop expressions
gives 5\% and 0.5\% accuracy for $C_7(\mw)$ and for
$C_9(\mw)$, respectively. 

In summary, we will compute  the QCD corrections to $Z^0$ penguin 
diagrams and to \ew boxes and exclude all the terms proportional 
to $\sws$ which are not quadratic in $\mt$.
Our approximation provides  the complete $\ord(\alpha_\smallw \as)$ corrections
to $C_{7-10}(\mw)$ not suppressed by $\sws$, as well as the full 
$\ord(\aw \as \sws \mt^2)$ correction. 
At the one-loop level, the combination of 
these two approximations reproduces very closely the full results.

\section{Renormalization Scheme Dependence} 
\label{RSdep}
\setcounter{equation}{0}
Next we would like to elaborate on the renormalization scheme
dependence of the Wilson coefficients and its cancelation in physical
amplitudes. For the purpose of our calculation we will need only the
transformation $\hat U (\mu, \mw, \aem)$ including LO and NLO
corrections and two NNLO terms to be specified below. 
Indeed as seen in (\ref{cI}) only $\hat U^{(0)}$ and $\hat
U^{(1)}$ enter $\vec C_I (\mu)$. At NLO we have
\cite{BJL93,ciuchini:94} 

\be\label{4.1}
\hat U (\mu, \mw, \aem) = \hat W (\mu) \hat 
U^{(0)} (\mu, \mw) \hat W^\prime (\mw) \, , 
\ee
where
\bea 
& & \hat W (\mu) = \Bigg(\hat 1 + \frac{\aem}{4 \pi} \hat J_{se} \Bigg)
\Bigg(\hat 1 + \frac{\as (\mu)}{4 \pi} \hat J_s \Bigg) \Bigg(\hat 1 +
\frac{\aem}{\as (\mu)} \hat J_e \Bigg) \, , \label{4.2}\\
& & \hat W^\prime (\mw) = \Bigg(\hat 1 - \frac{\aem}{\as (\mw)} \hat J_e \Bigg)
\Bigg(\hat 1 - \frac{\as (\mw)}{4 \pi} \hat J_s \Bigg) \Bigg(\hat 1 -
\frac{\aem}{4 \pi} \hat J_{se} \Bigg) \, . \label{wprime} 
\eea 
$\hat U^{(0)} (\mu, \mw)$ is the LO evolution matrix for which the
explicit expression can be found in \cite{buchalla:96,BJL93,ciuchini:94}. Also
expressions for $\hat J_s$, $\hat J_e$ and $\hat J_{se}$ can be found
there. They are functions of one-loop and two-loop anomalous
dimensions of the operators in question. 

From (\ref{4.1})--(\ref{wprime}) we extract 
\bea
\hat U ^{(1)}(\mu,\mw) & = & \frac{\as (\mu)}{4 \pi} \,\hat J_s \, \hat U^{(0)}
(\mu,\mw)-\frac{\as (\mw)}{4 \pi} \, \hat U^{(0)} (\mu,\mw) \hat J_s \, , 
\label{u1}\\
\hat R^{(0)}(\mu,\mw) & = & \frac{4 \pi}{\as (\mu)} \,\hat
J_e \, \hat U^{(0)}(\mu,\mw)
-\frac{4 \pi}{\as (\mw)} \, \hat U^{(0)} (\mu,\mw) \hat
J_e 
\eea
and suppressing the arguments of $\hat U^{(0)}$
\bea
\hat R^{(1)}(\mu,\mw)  &  =&  \hat J_{se} \,\hat U^{(0)} +
\hat J_s  \,\hat J_e \,\hat U^{(0)} - \frac{\as (\mu)}{\as(\mw)}  \,
\hat J_s \,\hat U^{(0)}  \,\hat J_e \non\\&-& \hat U^{(0)} \,\hat J_{se} +
\hat U^{(0)} \, \hat J_e  \,\hat J_s  - \frac{\as (\mw)}{\as(\mu)} 
 \,\hat J_e \,\hat U^{(0)} \, \hat J_s \, . 
\label{r1}
\eea
Now whereas $\hat r_s^{(1)}$, $\hat r_e^{(1)}$, $\hat J_s$ and $\hat
J_{se}$ depend on the renormalization scheme of operators it can be
shown \cite{BJL93} that
\be
\hat r_s^{(1)^T} + \hat J_s \, , \hspace{1.5cm} \hat r_e^{(1)^T} +
\hat J_{se}
\ee
are renormalization scheme independent. 
At NLO it follows \cite{BJL93} 
that the scheme dependence of $\hat U^{(0)} \vec C_s^{(1)}$ 
in (\ref{cs}) is canceled by the second term in $\hat U ^{(1)}$ multiplied by
$\vec C^{(0)}$. Similarly, the scheme dependences of $\hat U^{(0)} \vec
C_e^{(1)}$ and $\hat R^{(0)} \vec C_s^{(1)}$ in $\vec C_I$ and $\vec
C_{II}$ respectively are canceled by $\Delta \hat R^{(1)} \vec
C^{(0)}$ in (\ref{cII}), where $\Delta \hat R^{(1)}$ represents the three last
terms in (\ref{r1}). The remaining scheme dependences reside in 
$\Delta \hat U^{(1)} \vec C^{(0)}$ in (\ref{cs}) and $\Delta \hat
R^{(1)} \vec C^{(0)}$ in (\ref{cII}), where this time $\Delta \hat
R^{(1)}$  represents the first  
three terms  (\ref{r1}) and $\Delta \hat U^{(1)}$ the first term in 
(\ref{u1}). One can verify that the scheme dependence of these terms 
is canceled by the one of the matrix elements $\langle \vec Q(\mu)\rangle$.
To this end the matrix elements in (\ref{effective}) with $\ord(\as)$ and
$\ord(\alpha)$ terms retained and $\mw\to\mu$ should be used.

Turning to NNLO contributions in (\ref{cI}) let us concentrate on $\vec
C_{se}^{(2)}$ and in particular on the scheme dependent term in
(\ref{ces}) $\hat r_s^{(1)^T} \vec C_e^{(1)}$ which is taken into
account in our calculation. Keeping only this term in $\vec
C_{es}^{(2)}$ and adding the last term in (\ref{cI}) we obtain
\be \label{RSNNLO}
-\frac{\as (\mw)}{4 \pi} \, \hat U^{(0)}
\left (\hat r_s^{(1)^T} + \hat J_s \right) \vec C_e^{(1)} 
+\frac{\as (\mu)}{4 \pi} \, \hat J_s  \,\hat U^{(0)}  \,\vec C_e^{(1)}
\, . 
\ee
That is the scheme dependence of $\hat r_s^{(1)^T}$ in $\vec
C_{es}^{(2)}$ has been canceled by $\hat J_s$ in the second term in 
(\ref{u1}).
 However as $\vec C_e^{(1)}$ is renormalization
scheme dependent through $\hat r_e^{(1)^T}$ and $\hat J_s$ in the second term
in (\ref{RSNNLO}) is scheme dependent, both terms (\ref{RSNNLO}) remain scheme
dependent.  In order to cancel
the scheme dependence of the first term in (\ref{RSNNLO}) 
we would have to know other terms in
(\ref{ces}) which as discussed above we do not know. Fortunately at
NNLO the remaining scheme dependence in (\ref{RSNNLO}) does not bother
us as it is $\mt$-independent and the term $(\hat r_s^{(1)^T} \! + \hat
J_s) \, \hat r_e^{(1)^T}$ does not contribute to $\ord (\aw \as)$ and to
the contributions $\ord (\aw \as \sws)$ quadratic in $\mt$ considered
by us. Thus in evaluating the last term in (\ref{cI}) we will consistently
drop the terms originated in photon penguin diagrams, which are scheme
dependent. Applying this procedure also to the second term of
(\ref{RSNNLO}) one can easily verify that the remaining scheme
dependence of this term residing in $\hat J_s$
is canceled by the scheme dependence of $\langle \vec Q(\mu)\rangle$.
This procedure has to be properly implemented 
in our calculation of NNLO matching conditions. 
$\vec C_e^{(1)}$ in (\ref{RSNNLO}) is then simply given by
$C_{7-10} (\mw)$ with $\widetilde D_0 (x_t)$ removed. Needless to say
at NLO the full $\vec C_e^{(1)}$ should be included. Now in
Section \ref{res} we will perform the renormalization group evolution in
order to calculate $C_{7-10} (\mu)$ for $\mu \ll \mw$. From the
preceding discussion and (\ref{cI}) it should be clear that this
evolution should include the full NLO evolution modified by the
following NNLO terms:
\begin{itemize}
\item[i)] $\ord (\aw \as)$ and $\ord (\aw \as \sws \mt^2)$
contributions to $C_{7-10} (\mw)$.
\item[ii)] The contribution
\be \label{eq:ii}
\Delta \vec C (\mu) = \frac{\aem}{4 \pi} \left[ \frac{\as (\mu)}{4 \pi} \,
\hat J_s \,\hat U^{(0)} (\mu, \mw)
-\frac{\as (\mw)}{4 \pi} \,
\hat U^{(0)} (\mu, \mw) \hat J_s \right]\vec C_e^{(1)} 
\ee
representing the last term in (\ref{cI}) with 
$\vec C_e^{(1)}$ modified as discussed above.
It should be emphasized that (\ref{eq:ii}) is not included in the usual NLO
calculations as it is $\ord(\alpha\as)$ and belongs to the NNLO contributions.
\end{itemize}  

\section{QCD Corrections to the  $Z^0$-Penguin Diagrams} 
\label{Zpeng}
\setcounter{equation}{0}
The first part  of our analysis is devoted to the QCD corrections to the 
penguin diagrams originating in $Z^0$ exchange. It is convenient to
separate these corrections according to their structure in color space.
Let us denote by $\hat{1}$ and $T^a=\lambda^a/2$ the $N\times N$ matrices in
the color space of $SU(N)$. Since the relevant graphs always involve 
two quark lines,
the diagrams containing a gluon attached to a single quark line 
contribute to the color-singlet  component, characterized 
by $\hat{1} \otimes \hat{1}$, while diagrams where the gluon joins two
different quark lines contribute to the color-octet component, 
proportional to $T^a \otimes T^a$.

As far as  leading order and color-singlet two-loop
diagrams are concerned, the contributions of the $Z^0$-penguin vertex to the
$Q_i$  operators can be described in terms of an effective $\bar{s} d Z^0$
vertex
\be \label{sdzvertex}   
\Gamma_{\bar{s} d Z}^\mu = i \frac{g^3}{(16 \pi^2)} \f{\lambda_t}{\cw}    
        \,C (x_t) \, \bar{s} \gamma^\mu (1 - \gamma_5) d    
\ee  
with  $\lambda_t = V_{ts}^\ast V_{td}$ and $\cw=\cos\theta_\smallw$. 
The coefficient $C(x)$ at $\ord(\as)$ can
be written as 
\be \label{eqsecv:cfunction}   
C (x) = C_0 (x) + \frac{\as}{4\pi} \,C_1 (x) \, , 
\ee   
where $C_0(x) $, introduced in \equ{eq:C0},
is the relevant Inami-Lim function and  $C_1 (x)$, which was  
calculated in \cite{buchalla:93a,misiak99}, reads
\bea\label{C1}   
C_1 (x) & =& \f{29 x + 7 x^2 + 4 x^3}{3 (1 - x)^2}    
        - \f{x - 35 x^2 - 3 x^3 - 3 x^4}{3 (1 - x)^3} \ln x    
        - \f{20 x^2 - x^3 + x^4}{2 (1 - x)^3} \ln^2 x \non \\[1.5mm]  
        & & + \f{4 x + x^3}{(1 - x)^2} \mbox{Li}_2 (1 - x)    
        + 8 x \f{\partial C_0 (x)}{\partial x} \ln x_{\mu_t} \, .   
\eea   
Here we have used $x_{\mu_t} =  \mu_t^2/\mw^2$ and 
\be \label{polylog}
\mbox{Li}_2 (1 - x) =  
\int_{1}^{x} \! dt \, \frac{\ln t}{1 - t} \; , \hspace{5mm} x \geq 0 \, .  
\ee
The scale $\mu_t$ is the renormalization scale of the $\ms$ running 
top quark mass $\mt(\mu_t)$. We recall that $C_{0,1}$ depend on the 
gauge parameter of the $W$-field. This dependence is canceled at 
the level of Wilson coefficients by other contributions, to
be considered later on. \eqs{eq:C0} and (\ref{C1}) actually
hold in the 't~Hooft-Feynman gauge, $\xi_\smallw=1$. 

In terms of the effective $\Delta S=1$ Hamiltonian, the singlet 
 contribution of the $Z^0$-penguin
can be written as 
\bea \label{Heffz1}   
 \Heff^{\bar{s} d Z} = \f{G_F}{\sqrt 2} \f{\alpha_\smallw}{\pi   
     } \lambda_t \left[C_0(x_t)+\frac{\as}{4\pi} \, C_1 (x_t) \right] \sum_{q = u, d,    
        s, c, b} \! \Big [ \left (T_3^q - e_q \, s^2_\smallw\right) \cO_{LL}   
        - e_q \, s^2_\smallw \, \cO_{LR} \Big ] \, , \hspace{0.5cm}   
\eea   
where $T_3^q=\pm 1/2$ is the third component of the weak isospin,     
$e_q$ is the electric charge of the quark flavor $q$ and we have
introduced the shorthand notation $s_\smallw = \sin \theta_\smallw$.
The four-quark operators are given by
\bea      
\cO_{LL} &\!\!=\!\! & \bar{s} \, \gamma_\mu (1 - \gamma_5) \, d \; \bar{q} \,       
        \gamma^\mu (1 - \gamma_5) \, q  = (\bar s d)_{V-A}\;(\bar q       
        q)_{V-A} \, , \label{OLL} \\
\cO_{LR} &\!\!=\!\!& \bar{s} \, \gamma_\mu (1 - \gamma_5) \, d \; \bar{q} \,       
        \gamma^\mu (1 + \gamma_5) \, q  = (\bar s d)_{V-A}\;(\bar q       
        q)_{V+A} \, .            
\eea     
Using the identity $T_3^q = e_q - 1/6$, we can rewrite \equ{Heffz1}
in the basis  of the $Q_i$ operators
\bea \label{Heffz2}   
 \Heff^{\bar{s} d Z} = -\f{G_F}{\sqrt 2} \f{\alpha_\smallw}{6 \pi}  
   \lambda_t \left(C_0(x_t)+\f{\as}{4\pi}  C_1 (x_t) \right)
\Big [Q_3+4 \,s^2_\smallw \, Q_7   
        - 4\, c^2_\smallw \,Q_9 \Big ]    
\eea   
which modifies at $\ord(\as)$ the $C_0$ contributions to the Wilson coefficient
$C_i$ of (\ref{eq:CMw1})-(\ref{eq:CMw10}).

\begin{figure}[t]
\centerline{
\psfig{figure=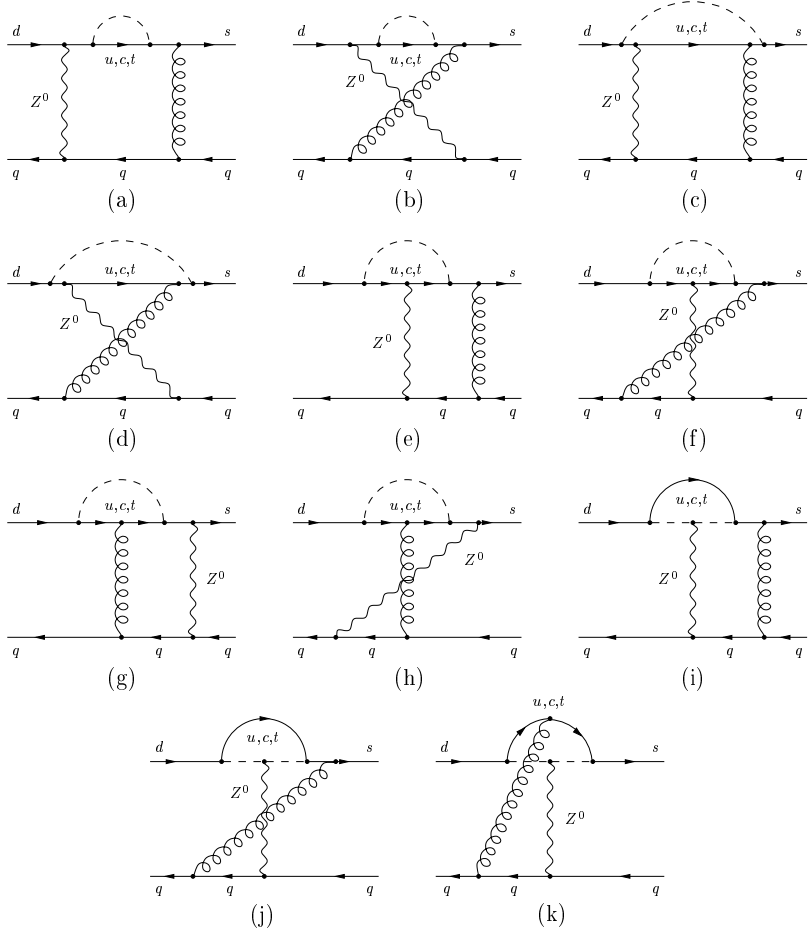,height=6.8in,rheight=4.7in}  }
\caption{\sf Feynman graphs contributing to the color-octet
  component of the $\ord(\as)$ corrections to  the $Z^0$ penguin vertex.
  Mirror diagrams are not displayed.}
\label{zzz}
\end{figure}
This completes the analysis of the color-singlet contribution.
We now proceed to the  calculation of the color-octet corrections to the
$Z^0$-penguin diagram, which is absent in the literature. 
We first calculate the two-loop
diagrams in the full theory. We will then compute the
renormalization contributions and finally match the renormalized amplitude of
the full theory with the result of the calculation of the $\ord(\as)$ corrections
to the effective theory as explained in Section 3.
The relevant two-loop SM diagrams are displayed in Fig.\,\ref{zzz}. 
It is important to realize that this is just a small 
subset of the $\ord(g^4 g_s^2)$ diagrams, which also include, for
instance, all the \ew corrections to the gluon penguin diagrams. Fortunately,
because of their flavor structure,  most  of them do not project 
on $Q_{7-10}$ and are not interesting for our purposes.
Only the diagrams involving a $Z^0$ or photon exchange across
the two quark lines\footnote{We recall that the \ew corrections to the flavor
conserving vertex of a gluon penguin diagram vanish as a result of Ward
identities \cite{MS80}.},
 as in Fig.\,\ref{zzz}, will contribute to the \ew penguin
operators, even if they are originated by a gluon penguin vertex
(Cf. Fig.\,\ref{zzz} (g)). 
According to the strategy elaborated in Section\,\ref{strategy},
we will compute only the $Z^0$ exchange diagrams.

All diagrams in Fig.\,\ref{zzz} except (c), (d) and (k) are infrared (IR) divergent. 
We regulate these divergences by the use of a common mass $m$ for the internal 
light quarks and set all external momenta to zero (Cf. \cite{buras:90}). 
It is noticeable that our results for the Wilson coefficients are
unchanged if we set all light quark masses to zero and regulate the 
IR-divergent Feynman integrals keeping a mass parameter only in the
denominators.  The IR 
divergences are canceled in the matching procedure by the contributions of the
effective theory.  

\begin{figure}[t]
\centerline{
\psfig{figure=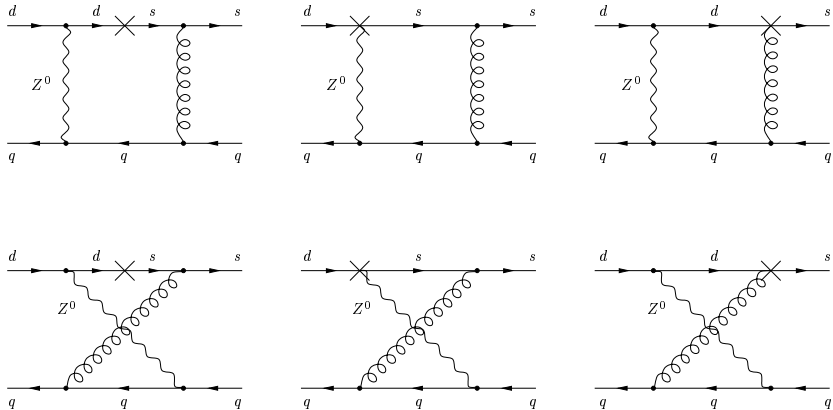,height=5.5in,rheight=2.3in}  }
\caption{\sf Counterterm diagrams for  the color-octet
  component of the $\ord(\as)$ corrections to the $Z^0$ penguin vertex.}
\label{ct}
\end{figure}

The IR divergent graphs have also ultraviolet (UV)
divergences which we regulate in $n=4- 2 \, \epsilon$ dimensions using an
anticommuting $\gamma_5$. 
Some of the UV divergences -- the ones
related to the exchange of a pseudo-Goldstone boson --
persist after implementation of the GIM mechanism. In a calculation of 
on-shell amplitudes, they would be canceled by external leg corrections.
However, the IR regularization we have adopted 
prevents the cancelations
among   the off-diagonal wave function renormalization of the internal quarks 
which are a prerequisite for this procedure. 
We are then forced to renormalize the amplitude at the diagrammatic level, 
and we do that by zero momentum subtraction  of the one-loop sub-divergences 
(the relevant counterterm diagrams are shown in Fig.\,\ref{ct}).
Specifically, writing the quark two-point function  for the $j\to i$
transition as 
\bea
\Sigma_{ij}(p) = 
\Sigma^{\smallL}_{ij}(p^2)  \not\!p \,P_{\smallL}+ 
\Sigma^{\smallR}_{ij} (p^2) \not\!p \,P_{\smallR} +
\Sigma^{\smallS}_{ij}(p^2) \left( m_i P_{\smallL}+m_j P_{\smallR} \right)
\label{2point} ,
\eea
where  $P_{\smallL,\smallR}=
\frac1{2} (1\mp \gamma_5)$ are
the left and right-handed  projectors and $m_{i,j}=m$, the subtraction involves
$\Sigma^{\smallL}_{ij}(0)$ and $\Sigma^{\smallS}_{ij}(0)$. 
$\Sigma^{\smallR}_{ij}(0)$ is $\ord(m^2)$ and can be neglected.
This subtraction procedure removes the spurious IR sensitivity of 
the diagrams in Fig.\,\ref{zzz} (a)-(b) and, in the limit $m\to 0$
we are considering,
implements the correct LSZ conditions on the external legs \cite{ckm}.
The case of the vertex-subdivergences is easier because 
the diagrams in Fig.\,\ref{zzz} (e)-(j) are less IR-sensitive.
One can therefore  neglect all terms proportional to $m$ in \equ{2point} 
  and the subtraction 
involves only $\not\!p \,P_{\smallL}\Sigma^{\smallL}_{ij}(0)$. 
For a more detailed discussion of the renormalization of off-diagonal 
quark amplitudes, see \cite{ckm,bbbar,aoki}.

We have performed two independent 
calculations, employing  a combination of  {\sc Mathematica} 
\cite{mathematica} routines  for the various stages of the computation,
 from the generation  of the Feynman diagrams
\cite{feynarts}, to the Dirac structure simplification \cite{tracer} 
and  the two-loop integration \cite{processdiag,tarcer}. 

After using the unitarity of the CKM matrix,
the renormalized two-loop amplitude in the full theory, suppressing the
external quark fields, can be written as
\be \label{M2loopzzz} 
M_{2loop} = \f{-i}{(16 \pi^2)^2} \f{g_s^2 g^4}{2 \mw^2} \lambda_t \sum_{k}  
        T^a \otimes T^a \, \cW_k^{\, (8)} (x_t, x_z) \, T_k \, ,   
\ee   
where the sum runs over $k = LL, LR, 1, 2$ and the spinor structures 
$T_k$ are given by 
\bea    \label{spinor}
T_{LL} & = & \gamma_\mu L \otimes \gamma^\mu L \, , \non \\   
T_{LR} & = & \gamma_\mu L \otimes \gamma^\mu R \, ,  \\         
T_1 \hspace{2.6mm} & = & L \otimes L + R \otimes L + L \otimes R +     
        R \otimes R \, , \non \\          
T_2 \hspace{2.6mm} & = & \sigma_{\mu \nu} \otimes \sigma^{\mu \nu},    \non\\
T_{3} \hspace{2.6mm} & = & \gamma_\mu R \otimes \gamma^\mu L +        
        \gamma_\mu L  \otimes \gamma^\mu R  \non            
\eea      
with $R, L = 1 \pm \gamma_5$. In order to project the renormalized
amplitudes on the  
different spinor structures, we use the method adopted for example 
in \cite{buras:93a} and reduce the
problem to the calculation of traces of strings of Dirac matrices.
As the amplitudes are now finite, this can be done in four dimensions.
The coefficients $\cW_k^{\, (8)} (x_t, x_z)$ can furthermore 
be decomposed according to
\bea   \label{W2loop}
\cW_{LL}^{\, (8)} (x_t, x_z) & = & \left ( T_3^q-e_q \, s^2_\smallw\right )    
        \Big [ \cZ (x_t, x_z) + 6 \, C_0 (x_t) \ln x_q \Big ] \, ,   
        \non\\
\cW_{LR}^{\, (8)} (x_t, x_z) & = & e_q \,s^2_\smallw\Big[\cZ(x_t,x_z)+ 6 \,    
  C_0 (x_t) \ln x_q \Big ] \, , \\   
\cW_{1}^{\, (8)} (x_t, x_z) & = &-(3+\xi)\,\left(T_3^q-2\,e_q\, s^2_\smallw\right)   
        C_0 (x_t) \, ,   \non\\
\cW_{2}^{\, (8)} (x_t, x_z) & = & (3+\xi) \, T_3^q \, C_0 (x_t) \, , \non 
\eea 
where $q$ is the flavor of the lower quark line (Cf. Fig.\,\ref{zzz}),
$x_z = \mz^2/\mw^2$
and the logs of $x_q=m^2/\mw^2$ indicate the IR divergences.  
In analogy to the case described in \cite{buras:90}, the structures
 $T_1$, $T_2$ and $T_3$ in \equ{M2loopzzz} are artefacts of the 
IR regularization procedure and we will verify in a moment that they 
drop out in the matching with the effective theory. For instance, if we 
consistently set the 
common quark mass to zero in the numerator of the quark propagator, 
$\cW^{\,(8)}_{1,2}$ vanish.
In \equ{W2loop} we have 
left the gluon gauge $\xi$ arbitrary and set $\xi_\smallw=1$. We have also
checked that the  $Z^0$-field gauge dependence of the individual diagrams 
 cancels in their sum.

\begin{figure}[t]
\centerline{
\psfig{figure=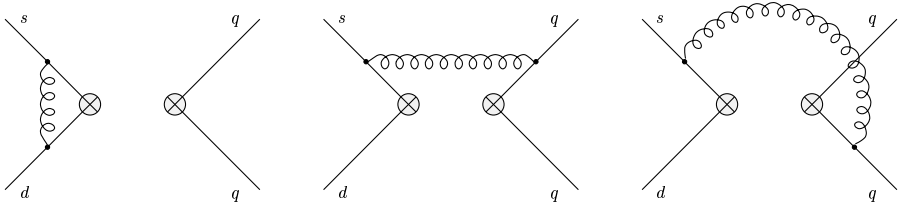,height=5.7in,rheight=1.4in}  }
\caption{\sf Diagrams contributing to the matrix element of current-current
operators at $\ord(\as)$.}
\label{eff}
\end{figure}

For what concerns the effective theory side, we need the octet-part 
of the one-loop matrix elements of the renormalized 
operators $\cO_{LL}$ and $\cO_{LR}$ 
in QCD. The calculation is performed following the same 
regularization procedure used for the   two-loop diagrams and it involves the
one-loop diagrams depicted in Fig.\,\ref{eff}. In principle also
insertions in the penguin diagrams should be considered. However at
the level of the approximations outlined in Section 3 they do not
contribute to the Wilson coefficients of $Q_{7-10}$. 
After renormalization in the NDR scheme, we obtain
\bea \label{OLLoneloop}      
\langle \cO_{\! LL} (\mu) \rangle_{\rm 1 loop} &=& \langle \cO_{\! LL}         
        \rangle_{\rm tree} + \f{\as(\mu)}{4\pi} 
        \sum_{k} \! \left [ C_F \, \hat{1} \!         
      \otimes \! \hat{1} \, \cA^{(1)}_k (\mu) + T^a \! \otimes \! T^a     
     \cA^{(8)}_k (\mu) \right ] \, T_k \, ,   
        \hspace{0.5cm}   \\
\label{OLRoneloop}      
\langle \cO_{\! LR} (\mu) \rangle_{\rm 1 loop} &=& \langle \cO_{\! LR}        
        \rangle_{\rm tree} + \f{\as(\mu)}{4\pi} \sum_{k} \! \left [ C_F \, \hat{1} \!      
      \otimes \! \hat{1} \, \cB^{(1)}_k (\mu) + T^a \! \otimes \! T^a          
      \cB^{(8)}_k (\mu) \right ] \, T_k 
        \hspace{0.5cm} 
\eea       
with $k = LL,LR, 1,2,3$. The coefficients 
$\cA_k$ and $\cB_k$ which do not vanish   are given by
\bea  \label{Ai}         
\cA^{(1)}_{LL} (\mu) & = & -3 - 2 \, \xi \ln x_q + 2 \, \xi \ln \xmu \non     
        \, , \\          
\cA^{(8)}_{LL} (\mu) & = & -5 + 6 \ln x_q - 6 \ln \xmu \, , \\ 
\cA^{(8)}_1 (\mu) & = & -\cA^{(8)}_2 (\mu)=-2 \, \cA^{(1)}_3 (\mu)\ = \ 
-(3 + \xi) , \non\\        
\cB^{(1)}_{LR} (\mu) & = & -3 - 2 \, \xi \ln x_q + 2 \, \xi \ln \xmu        
        \, , \non\\         \label{Bi}
\cB^{(8)}_{LR} (\mu) & = & -7 - 6 \ln x_q + 6 \ln \xmu \, , \\
\cB^{(8)}_1 (\mu) & = & \cB^{(8)}_2 (\mu) \ = \
       - 2 \, \cB^{(1)}_3 (\mu) \ =  \ -(3 + \xi)
\non\, .   
\eea          
The results for $\cA_k$  can be also obtained from \cite{buras:90},
after taking the limit $m_s=m_d=m$.  

Unlike the full theory results,
(\ref{Ai})-(\ref{Bi}) are scheme dependent. For instance the
constant terms  depend on the way  
$\gamma_5$ is defined in $n$ dimensions --- in our case
 they are specific to the
NDR scheme. 
The scheme dependence is generated in the calculation of the 
matrix elements in the effective theory:
for example, in the Dimensional Reduction 
(DRED) scheme \cite{siegel,buras:90c} there is no constant part in 
$\cA^{(i)}_k$ and $\cB^{(i)}_k$ but only logarithms. In the 
 't~Hooft-Veltman (HV) scheme \cite{HV}, the constants in
$\cA^{(1,8)}_{LL}(\mu)$ are (1, $-1$) instead of $(-3, -5)$, 
respectively, and in $\cB^{(1,8)}_{LR}(\mu)$ they 
are (1, 5) instead of $(-3,-7)$ (see also (3.9)-(3.10) of \cite{buras:93a}).
(\ref{Ai})-(\ref{Bi}) also depend on the definition of 
the evanescent operators \cite{dugan,herrlich:95}. 
It is crucial that this  definition 
 follows the one adopted in the calculation of the two-loop
$\ord(\as^2)$ anomalous dimension matrix \cite{buras:93a,ciuchini:94}. In
practice, in our case this means  that we have  to perform
the projection on $T_k$
by taking traces in $n$ dimensions with anticommuting $\gamma_5$.
As a consequence of our choice of IR regularization,
and in contrast to \cite{misiak99,buras:99a},
this is the only occurrence of the evanescent operators in our calculation.


We have now all ingredients needed to match full and effective theories.
Taking into account  \eqs{M2loopzzz}, (\ref{OLLoneloop}) and (\ref{OLRoneloop}),  we see 
that the color-octet part of the effective Hamiltonian can be written as 
\bea \label{HeffZg1} 
 \Heff^{Z,  8} & = & \f{G_F}{\sqrt 2} \f{\alpha_\smallw}{\pi }
\f{\as}{4\pi}  \lambda_t
        \, T^a \otimes T^a \! \! \! \! \sum_{q = u, d, s, c, b} \! 
      \Big \{ \left (T_3^q - e_q \, s^2_\smallw \right ) G \,
        \cO_{LL}  + \, e_q \, s^2_\smallw H \,\cO_{LR} \Big \} ,
\eea   
where the coefficients $G $ and $H $ are given by 
\bea \label{G} 
G  & \equiv &    
        \cZ (x_t, x_z, s_\smallw) + 5 \, C_0 (x_t)   
        + 6 \, C_0 (x_t) \ln x_{\mu_W} \, , \\[1mm]   
H         & \equiv & \cZ (x_t, x_z, s_\smallw) - 7 \, C_0 (x_t)   
        + 6 \, C_0 (x_t) \ln x_{\mu_\smallw} \label{H} 
\eea  
and we have introduced $x_{\mu_W}=\mu_\smallw^2/\mw^2$.
The unphysical and $\xi$-dependent 
terms obtained from the two-loop calculation of \equ{M2loopzzz} have been 
canceled by analogous terms from the effective theory. Note that 
the scale $\mu_\smallw$ 
in (\ref{G})-(\ref{H}) is the scale at which the matching
is performed. The scale $\mu_\smallw$ is not related to 
the top quark mass renormalization scale $\mu_t$ appearing in 
\equ{C1} although they can be set equal. We will, however,  keep them
distinct in the following. Taking advantage  of 
the identity $T_3^q = e_q - 1/6$ and using 
\bea
T^a \otimes T^a = \f{1}{2} \left (\tilde{1} \otimes \tilde{1}           
        - \f{1}{N} \hat{1} \otimes \hat{1} \right) \,       , \label{colorid}
\eea
where $\tilde{1}$ indicates twisted color indices -- 
like in $Q_4$ of  \equ{OS2} --
we can rewrite \equ{HeffZg1} in terms of the $Q_i $ operators:
\be  \label{finZ}
 \Heff^{Z,8}  =    
    \f{G_F}{\sqrt 2} \f{\alpha_\smallw}{36 \pi} \f{\as}{4\pi} \,\lambda_t   
        \left[ G   \left(Q_3-3\,Q_4\right)   
        - 4\,  s^2_\smallw H \left( Q_7-3\,Q_8\right)   
 - 4 \,c^2_\smallw \,G  \left(Q_9-3 \,Q_{10}\right)    
        \right] .  
\ee  
From here one can read  the contributions of this class of diagrams 
to the various  Wilson coefficients at the matching scale $\mu_\smallw$ 
calculated for  $\xi_\smallw=1$  in the NDR scheme.

As we have seen above, 
at the one-loop level and in the case of the color-singlet $\ord(\as)$ corrections
the dependence on $\sin^2\theta_W$ drops out in the functions  $C_{0,1}(x_t)$.
This is a consequence of the Ward identity which ensures that the photon 
exchange diagram has no $1/q^2$ pole, and it is guaranteed because the momentum
carried by the $Z^0$ boson is vanishingly small. 
In the case of the octet contributions the Ward identity 
does not  hold, because the momentum carried by the $Z^0$ is not small. Indeed,
 we verify that the $\sin^2\theta_W$-dependence is not removed from 
the Wilson coefficient and that the function $\cZ$ can be decomposed into   
\be \label{decomp}
\cZ (x_t, x_z, s_\smallw) = \cZ_0 (x_t, x_z) + s^2_\smallw \, \cZ_1 (x_t, x_z).   
\ee     
The coefficients $\cZ_{0,1}$  are  complicated functions of $x_t$ and $x_z$
and are given in \eqs{Z0} and (\ref{Z1}) of the Appendix.
As $\mt$ and $\mw$ are now accurately determined,
$\cZ_{0,1}$  can be linearized in the vicinity of their 
central values. Using the  latest experimental 
results  $\mt(\mt)=166\pm 5$ GeV,  $\mw=80.394\pm 0.042$ GeV,
and $\mz=91.1867$ GeV \cite{LEPEWWG}, we find
\bea\label{linearized}
{\cal Z}_0&=&+5.1795 + 0.038 \,(\mt-166 ) + 0.015 \,(\mw-80.394),
\non\\
{\cal Z}_1&=&-2.1095 + 0.0067 \,(\mt-166 ) + 0.026 \,(\mw-80.394 )
\eea
which reproduce the analytic expressions to great accuracy,  better than 
0.1\%, within 2$\sigma$ from the central values. 

It is also interesting to see how the HTE
approximates these two functions. In this respect we stress that, 
although $G$ and $H$ are $\xi_\smallw$-dependent quantities, 
their leading HTE term is gauge-independent, as
$Z^0$-penguins are the only source of contributions quadratic 
in $\mt$. The contributions quadratic in $\mt$ are 
\be \label{Zhte}
\cZ_0^{HTE}=  \frac{3}{4} \, x_t \left(1-  \ln x_z \right) \, ,
\hspace{1cm} \cZ_1^{HTE}=0 
\ee
so that we find  $\cZ_0^{HTE}=2.39$ and $\cZ_1^{HTE}=0$  
at leading order in the HTE.
At next-to-leading order in the HTE the approximation improves substantially, as we get 
$\cZ_0=4.62$ and $\cZ_1= -1.46$, 
relatively close to the central values of 
\equ{linearized}. Finally, we  notice
 that the leading term of the HTE can be  obtained considering  only 
the diagrams involving Yukawa couplings of the top quark, as we have explicitly
verified. 

In summary, in this section we have calculated the gluonic corrections
to the $Z^0$-penguin diagrams. The main results are (\ref{Heffz2}) and
(\ref{finZ}). 

\section{QCD Corrections to the Electroweak Box Diagrams}
\label{boxes}
\setcounter{equation}{0}
The second part of our analysis concerns the \ew box diagrams. Again, 
we will consider for definiteness the case of $\Delta S=1$ transitions. 
Although  some $\ord(\as)$ results are available in the literature
 for the case in which all quark involved in the
transition are down quarks \cite{buras:90,urban}
and for the case of semi-leptonic transitions 
\cite{buchalla:93b,misiak99,buras:99a},
the \ew box diagrams involving both down and up quark lines require a
new calculation that we describe in this section. 
Indeed, it is a fortuitous coincidence that at the one-loop level
quark box diagrams  
containing either up or down quarks are described by the single 
function $B_0(x_t)$ introduced in \equ{eq:B0}.
As a by-product of this computation we will also be able to reproduce all the 
two-loop box results of \cite{buras:90,urban,misiak99,buras:99a}.

First, we need to recall some one-loop results necessary for 
the subsequent discussion. The one-loop amplitude for
$\bar{s} + d \to \bar{q} + q$ with   $q = u, c$ can be written as 
\be \label{m1loopup1}           
M_{1loop} = \f{-i}{16 \pi^2} \f{g^4}{4 \mw^2} \sum_{i, j}        
        \lambda^{(q)}_i \lambda_j \, \cS^{(u)} (x_i, x_j) \cO_{LL} ,
\ee           
where $\lambda^{(q)}_i = |V_{qi}|^2$,  $\lambda_j = V_{js}^{\ast}        
V_{jd}$ and $\cO_{LL}$ has been defined in \equ{OLL}.
The function $ \cS^{( u)} (x_i, x_j)$ describes a generic $\Delta S=1$ box 
with external up quarks and  arbitrary internal quark masses $m_{i,j}$.
Expanding it up to $\ord(\epsilon)$, it reads 
\be \label{Su}           
\cS^{( u)} (x_i, x_j) = \cS^{( u)}_0 (x_i, x_j) + \epsilon \,        
        \cS^{( u)}_1  (x_i, x_j, \xmu) + \cO (\epsilon^2)           
\ee           
with 
\bea            
 \label{Su0}& & \hspace{-1.8cm} \cS^{( u)}_0 (x_i, x_j) \hspace{2mm} =  
         \hspace{2mm} -\f{16 - 7 x_i x_j}{16 (x_i - 1) (x_j - 1)}         
        - \left [\f{x_i^2 (16 + x_j (x_i - 8))}{16 (x_i - 1)^2 (x_i -        
        x_j)}  \ln x_i + (x_i \longleftrightarrow x_j) \right ] \, ,        
        \\  \label{Su1}        
& & \hspace{-1.8cm} \cS^{( u)}_1 (x_i, x_j, \xmu) \hspace{2mm} =         
        \hspace{2mm} -\f{40 - 13 x_i x_j}{32 (x_i - 1) (x_j - 1)}            
        -\Bigg [ \f{x_i^2 (40 + x_j (3 x_i - 16))}{32 (x_i - 1)^2 (x_i        
        - x_j)}  \ln x_i \non \\ & & \hspace{1.7cm} -\f{x_i^2 (16        
        + x_j (x_i - 8))}{32 (x_i - 1)^2  (x_i - x_j)} \ln^2 x_i +        
        (x_i \longleftrightarrow x_j) \Bigg ] + \cS^{(\rm u)}_0 (x_i,        
        x_j) \ln \xmu            
\eea           
and $x_{i,j} = m_{i,j}^2/\mw^2$. $\cS^{( u)}$ depends on the $W$-field gauge
and the above expressions hold in the 't~Hooft-Feynman gauge.
Setting all light quark masses to zero and using the unitarity of the CKM 
matrix, the only relevant combination in the limit $\epsilon \to 0$ is 
\be           \label{boxup}
\cS^{( u)} (x_t, 0) - \cS^{( u)} (0, 0)= - 4 \, B_0(x_t) \, ,
\ee      
where $B_0(x_t)$ is the Inami-Lim function of (\ref{eq:B0}).
Taking advantage of the identity $\sum_{q = u, c} \cO_{LL} = \f{1}{3} Q_3 +
 \f{2}{3} Q_9 $, we obtain the effective Hamiltonian induced by the
box diagrams with
isospin $T_3=1/2$ (up) quarks:
\be \label{Heffu0}           
\Heff (T_3 = 1/2) = -{G_F \over \sqrt{2}} \f{2\alpha_\smallw}{3 \pi      
        } \lambda_t \, B_0 (x_t) \left ( Q_3 + 2 \, Q_9        
        \right) \, .            
\ee           

The case of the down-quark box diagrams is slightly more complicated in that there is 
a mismatch in the CKM factor between the $q=d,s$ and $q=b$ cases. 
This implies the introduction of two additional operators
$Q_{11,12}$ \cite{buras:90b}
\be           
Q_{11} = (\bar s d)_{V-A}\;(\bar b b)_{V-A} \, , \hspace{1cm} Q_{12} =         
        (\bar s b)_{V-A}\;(\bar b d)_{V-A} \, .           
\ee           
Calling $\cS^{(d)}(x_i, x_j)$ the box function for the down-quark box diagrams,
we find after GIM
\bea \label{down2}           
M_{1loop}^{(d, s)} = \f{-i}{16 \pi^2} \f{g^4}{2 \mw^2} \Big \{           
        \lambda_t \left[\cS^{(\rm d)}(x_t, 0) - \cS^{(\rm d)}(0, 0)        
        \right] \Big \} \; \cO_{LL} \, ,  
\eea  
where we have dropped a term suppressed by $\lambda^{(d,s)}_t$.
On the other hand,   in the case of $b$ quarks $\lambda^{(b)}_t$ is not a
suppression factor and we obtain in this case
\bea \label{bottom2}           
M_{1loop}^{(b)} = \f{-i}{16 \pi^2} \f{g^4}{4 \mw^2} \Big \{           
        \lambda_t \left[\cS^{( d)}(x_t, 0) - \cS^{( d)}(0, 0)        
        \right]  \cO_{LL} \hspace{5.8cm} \non \\           
        + \; \lambda^{(b)}_t \lambda_t \left [\cS^{( d)}(x_t, x_t)         
        - 2 \, \cS^{(d)}(x_t, 0) + \cS^{( d)}(0, 0) \right]           
        \big ( Q_{11} + Q_{12} \big ) \Big \} \, , \hspace{0.5cm}           
\eea           
The function $\cS^{( d)}(x_i,x_j) $ undergoes the same decomposition of
\equ{Su}. In the 't~Hooft-Feynman gauge the  coefficients take the form
\bea           
& & \hspace{-2cm} \cS^{( d)}_0 (x_i, x_j) = \f{4 - 7 x_i x_j}{16        
        (x_i - 1) (x_j - 1)} + \left[ \f{x_i^2 (4 + x_j (x_i - 8))}{16        
        (x_i - 1)^2 (x_i - x_j)} \ln x_i + (x_i \longleftrightarrow        
        x_j) \right]  \, , \\ \label{Sd1}
& & \hspace{-2cm} \cS^{( d)}_1 (x_i, x_j, \xmu) =           
        -\f{4 + 13 x_i x_j}{32 (x_i - 1) (x_j - 1)}            
        + \Bigg [ \f{x_i^2 (-4 + x_j (3 x_i - 16))}{32 (x_i - 1)^2        
        (x_i - x_j)} \ln x_i \non 
\\  & & \hspace{1.1cm} -        
        \f{x_i^2 (4 +  x_j (x_i - 8))}{32 (x_i - 1)^2 (x_i - x_j)}        
        \ln^2 x_i + (x_i \longleftrightarrow x_j) \Bigg ] + \cS^{(        
        d)}_0 (x_i, x_j) \ln \xmu \, .            
\eea           
In $n=4 $ dimensions the combinations present in \eqs{down2} and (\ref{bottom2})
 reduce to
\be          \label{boxdown} 
 \cS^{( d)}(x_t, 0) - \cS^{( d)}(0, 0)= B_0(x_t)           
\ee
and     
\bea         
\cS^{(d)}(x_t, x_t) - 2  \cS^{( d)}(x_t, 0) + \cS^{( d)}(0, 0)
\equiv {1 \over 4} S_0 (x_t)             \label{Bd0}           
 =  \f{1}{4} \! \left [ \f{4 x_t \! - \! 11 x_t^2 \! + \!        
        x_t^3}{4 (1 \! - \!  x_t)^2} \!  - \! \f{3 x_t^2 \ln x_t}{2 (1
\! - \! x_t)^3}        
        \right ] , \hspace{0.5cm}      
\eea          
where  $S_0 (x_t)$ is the box function characteristic of $\Delta F=2$
transitions.  
We see from \eqs{boxup} and (\ref{boxdown}) that $T_3=1/2$ and 
$T_3=-1/2$ box diagrams involve the same function $B_0(x_t)$. This is true only
in $n=4$ dimensions and for the 't~Hooft-Feynman gauge. We will see in the
following that there is no such relation at $\ord(\as)$.

Using the identity $\sum_{q = d, s, b} \cO_{LL} = \f{2}{3} (Q_3 -  Q_9) $,
we can write the contribution to the effective Hamiltonian as  
\bea \label{Heffd01}          
\Heff (T_3 = -1/2) = {G_F \over \sqrt{2}} \f{\alpha_\smallw}{2 \pi        
        } \lambda_t 
\Bigg \{ \f{2}{3}        
        B_0 (x_t) \left ( Q_3 - Q_9 \right) + \f{1}{4} \lambda^{(b)}_t        
        \, S_0 (x_t) \left (Q_{11} + Q_{12} \right) \Bigg \} .
\eea            
The role of the operators $Q_{11,12}$  in the RGE evolution of the Wilson
coefficients between $\mw$ and $m_b$  has been analyzed in \cite{buras:90b}.
In the case of $\varepsilon'/\varepsilon $, for instance, 
 they can be  safely neglected. Their $\ord(\as)$ corrections are likely to be 
irrelevant and will not be considered in the following.

\begin{figure}[t]
\centerline{
\psfig{figure=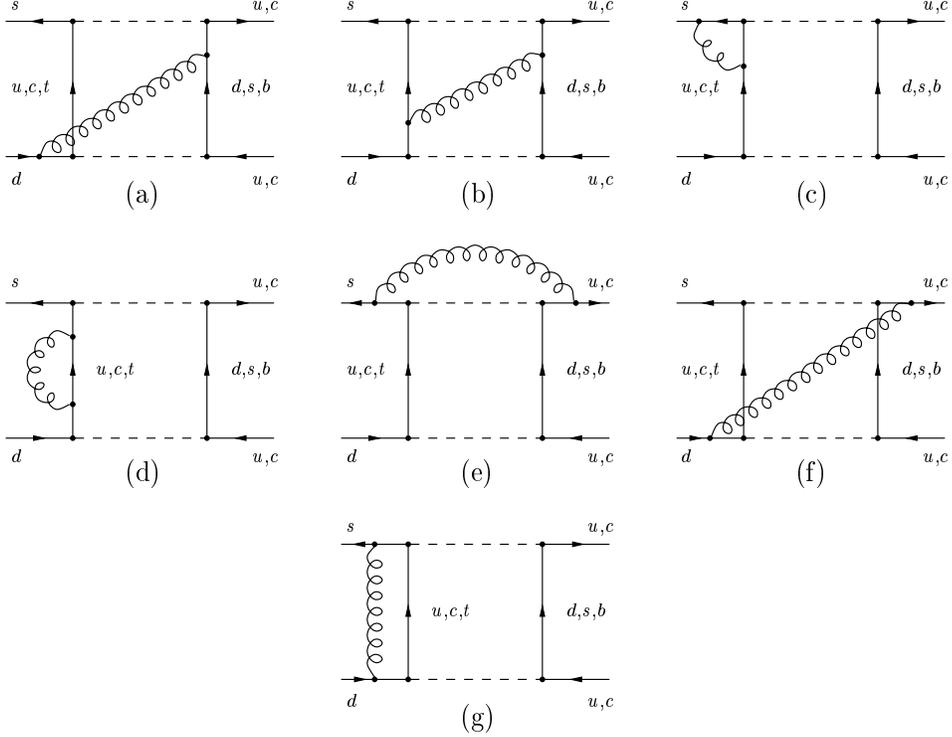,height=6.9in,rheight=3.6in}  }
\caption{\sf 
 Feynman graphs contributing to
 the $\ord(\as)$ corrections to the \ew box diagrams.
  Mirror diagrams are not displayed.}
\label{box}
\end{figure}
We are now in the position to present the calculation of the gluonic
corrections to the one-loop \ew box diagrams. The relevant diagrams for the case 
of isospin $T_3=1/2$ are shown in Fig.\,\ref{box}. Both color-singlet
(c), (d), (g) 
and color-octet (a), (b), (e), (f) diagrams are present.
The calculation proceeds
along the same lines as the one of Section\,\ref{Zpeng}. Diagrams (e),
(f) and (g) present IR
divergences which are regulated in the way described in the previous section.
The origin and the treatment of the UV divergences, however, is 
fundamentally different: diagrams (c), (d) have subdivergences related to the 
quark-gluon interactions and are renormalized in the $\ms$ scheme.
In fact, it is sufficient to renormalize the internal  quark masses
and to implement the wave function renormalization of the external legs.
Of course, in the counterterm diagrams the $\ord(\epsilon)$ parts of
(\ref{Su1}) and (\ref{Sd1}) have to be retained. 
The renormalized amplitude for the process $\bar{s} + d        
\to \bar{q} + q$ with $q = u, c$ can be written as 
\be \label{M2loopup}           
M_{2loop}^{box} = \f{-i}{(16 \pi^2)^2} \f{g_s^2 g^4}{4 \mw^2}            
        \sum_{i, j} \lambda^{(q)}_i \lambda_j \; 
\sum_{k} \left [ C_F \, \hat{1} \otimes        
        \hat{1} \, \cV^{\, ({u}, \, 1)}_k  + T^a \otimes        
        T^a \, \cV^{\, (u, \, 8)}_k  \right] T_k           
\ee           
with  $k = LL, 1,2,3$. $N$ is the number of colors and  
$C_F =(N^2 - 1)/2 N$. 
The spinor structures $T_k$ 
have been introduced in \equ{spinor}. Setting $\xi_\smallw=1$,
  keeping the gluon gauge parameter $\xi$ arbitrary, and 
without making assumptions on the masses of the internal quarks, we find
\bea \label{cVup1}           
\cV^{\, ({u}, \, 1)}_{LL} (x_i, x_j) & = & \cL^{\, ({u}, \, 1)}        
        (x_i, x_j) - 2 \, \xi \ln x_q \, \cS^{( u)}_0 (x_i, x_j)        
         \\ 
& & + 2 \, \xi \ln \xmu \, \cS^{( u)}_0 (x_i,         
        x_j) + 6 \ln \xmu \left (x_i \f{\partial}{\partial x_i}            
        + x_j \f{\partial}{\partial x_j} \right) \cS^{( u)}_0 (x_i,        
        x_j) \, , \hspace{1cm} \non\\
\label{cVup2}         
\cV^{\, ({u}, \, 8)}_{LL} (x_i, x_j) & = & \cL^{\, ({u}, \, 8)}         
        (x_i, x_j) + 6 \ln x_q \, \cS^{( u)}_0 (x_i, x_j) \, , \\          
\label{cVup3}         
\cV^{\, ({u}, \, 8)}_1 (x_i, x_j) & = & -
\cV^{\, ({u}, \, 8)}_2 (x_i, x_j)=-2 \, \cV^{\, ({u}, \,        
        1)}_3 (x_i, x_j) =  -(3 + \xi) \, \cS^{( u)}_0 (x_i, x_j) \, .
\eea            
All remaining $\cV_k^{(u,i)}$ vanish.
The two terms in the second line of \equ{cVup1} describe the scale dependence
introduced by the $\ms$ renormalization of the external fields and of the 
internal masses, respectively.  
The functions $\cL^{(u,i)}$ are  independent  of the gluon gauge. The
complete expressions are quite long and can be found in \cite{uli}. 

Concerning the effective theory, only the insertion of the  
operator $\cO_{LL}$ is relevant in this case and the results can be found in
the previous section. It is easy to verify that all the unphysical 
spinor structures and the 
gauge-dependent terms of (\ref{cVup1})-(\ref{cVup3})
cancel in the matching and we are left only with contributions 
proportional to $T_{LL}$. Using the unitarity of the CKM matrix
and  \equ{colorid},
the matching of full and effective theory 
leads to the following contribution to the effective Hamiltonian
\bea           \label{finBu}
\Delta \, \Heff(T_3 = 1/2)  =                  
        {G_F \over \sqrt{2}} \f{\alpha_\smallw}{6 \pi} \f{\as}{4\pi} 
        \lambda_t        
 \left [ B^{ u}_1 (x_t)    
        \left (Q_3 + 2 \, Q_9 \right) + \widetilde{B}^{ u}_1 (x_t)   
        \left (Q_4 + 2 \, Q_{10} \right) \right ] .
\eea           
The functions  $B^{ u}_1 (x_t)$ and        
$\widetilde{B}^{ u}_1 (x_t)$  are given by 
\bea \label{Bu1}        
B^{u}_1 (x_t) & = &  
-\f{2 x_t (23 + 9 x_t)}{3 (x_t -       
        1)^2} - \f{16 x_t (1 - 5 x_t)}{3 (x_t - 1)^3} \ln x_t            
        - \f{x_t (9 + 23 x_t)}{2 (x_t - 1)^3} \ln^2 x_t  \\
   & & - \f{6 x_t}{(x_t - 1)^2} \mbox{Li}_2 (1 - x_t) 
       -\f{38}{3} B_0 (x_t) + 4 B_0 (x_t) \ln x_{\mu_W}         
        -32 x_t \f{\partial B_0 (x_t)}{\partial x_t}        
         \,\ln x_{\mu_t} , \non \\
\label{Bu1t}    
\widetilde{B}^{ u}_1 (x_t) & = & 
 -\f{6 x_t}{x_t - 1}            
        - \f{3 x_t}{2 (x_t - 1)^2} \ln^2 x_t            
        - \f{6 x_t}{(x_t - 1)^2} \mbox{Li}_2 (1 - x_t)
\non \\ &  & -  10 \,B_0 (x_t)            
         -12\, B_0 (x_t) \ln x_{\mu_W} \, . 
\eea           
Again, these results hold in the 't~Hooft-Feynman gauge $\xi_\smallw=1$ and
are specific to the NDR scheme. As we will discuss in more detail later,
the scheme dependence resides in the coefficients $-38/3$ and $-10$ that
multiply $B_0(x_t)$.
Notice also that $B^u_1 $ depends on both $\mu_t $ and $\mu_\smallw$.

We now consider the case of $T_3=-1/2$ box diagrams. The relevant 
two-loop diagrams are the analogue of 
the ones shown in Fig.\,\ref{box}, although
in this case one should also consider the Fierz rotated diagrams, which 
just lead to an overall factor of 2. The
renormalized amplitude can therefore be written in the same way 
as in \equ{M2loopup}, but it is 
characterized by new coefficients $\cV^{(d,i)}$. These coefficients agree 
with the expressions given in the Appendix of \cite{buras:90}.
After the matching with the effective theory and the implementation of the GIM
mechanism, we can express the contribution to the effective Hamiltonian 
of the weak isospin $T_3 = -1/2$ box diagrams as      
\bea           \label{finBd}
\Delta \, \Heff(T_3 = -1/2) =                 
        {G_F \over \sqrt{2}} 
\f{\alpha_\smallw}{3 \pi } \f{\as}{4\pi} \lambda_t  \left [ B^{ d}_1       
        (x_t) \left ( Q_3 -  Q_9 \right) + 
        \widetilde{B}^{d}_1 (x_t) \left (Q_4 -   Q_{10}    
        \right) \right ] ,
\eea           
where the functions $B^{ d}_1$ and $\widetilde{B}^{d}_1 (x_t)$
are given by
\bea \label{Bd1}        
B^{ d}_1 (x_t) \!\! & = & \!\! 
-\f{8 - 183 x_t + 47 x_t^2}{24 (x_t -       
        1)^2} - \f{8 + 27 x_t + 93 x_t^2}{24 (x_t - 1)^3} \ln x_t +       
        \f{x_t (27 + 71 x_t - 2 x_t^2)}{24 (x_t - 1)^3} \ln^2 x_t \non\\       
& & - \f{2 - 3 x_t - 9 x_t^2 + x_t^3}{6 x_t (x_t -       
        1)^2} \mbox{Li}_2 (1 - x_t) + \f{2 + x_t}{6 x_t} \, \zeta (2) 
     + \f{19}{6} B_0 (x_t) \non\\
&& - B_0(x_t) \ln x_{\mu_W} +  8 x_t 
        \f{\partial B_0 (x_t)}{\partial x_t} \, \ln x_{\mu_t} \, , \\     
\label{Bd1t}      
\widetilde{B}^{d}_1 (x_t)\!\!      & = &\!\!        
  -\f{8 - 23 x_t}{8 (x_t - 1)}            
        - \f{8 - 5 x_t}{8 (x_t - 1)^2} \ln x_t            
        + \f{x_t (3 + 2 x_t)}{8 (x_t - 1)^2} \ln^2 x_t - \f{2 + x_t}{2 x_t} 
\, \zeta (2)\non\\ 
&&        + \f{2 - 3 x_t + 3 x_t^2 + x_t^3}{2 x_t (x_t - 1)^2}       
        \mbox{Li}_2 (1 - x_t) 
         + \f{5}{2} B_0(x_t) + 3 B_0 (x_t) \, \ln x_{\mu_W} \, .   
\eea           
As before, 
the previous expressions are specific to the NDR scheme and are valid for
$\xi_\smallw=1$. 

Summarizing, in this section we have calculated the NNLO contributions
originated in electroweak box diagrams. The main results are reported
in (\ref{finBu}) and (\ref{finBd}).

\section{Numerical Results}
\label{res}
\setcounter{equation}{0}
In this section we summarize our results in terms of $\ord(\alpha_\smallw\as)$
contributions to the Wilson coefficients of the \ew penguin operators 
$Q_{7-10}$ and study their numerical relevance, both at the \ew scale and
at typical hadronic scales in the NDR and HV schemes. We discuss the
reduction of the $\mu_t$-dependence in the Wilson coefficients and the
issue of the renormalization scheme dependence. 
We conclude with a discussion of the universality of the functions $X$ and $Y$
of the Penguin-Box Expansion.

\subsection{Results for the Wilson Coefficients}
Let us   collect the results of (\ref{Heffz2}), (\ref{finZ}),
(\ref{finBu}) and (\ref{finBd}). Using  
\bea 
G_0  (x_t, x_z) & = &    
        \cZ_0 (x_t, x_z) + 5 \, C_0 (x_t)   
        + 6 \, C_0 (x_t) \ln x_{\mu_W} \, , \label{G0}  \\ 
H_0  (x_t, x_z) & = & \cZ_0 (x_t, x_z) - 7 \, C_0 (x_t)   
        + 6 \, C_0 (x_t) \ln x_{\mu_\smallw} \, ,
\label{H0} 
\eea  
we obtain the following $\ord(\alpha_\smallw \as)$ corrections
to the Wilson coefficients of the electroweak penguin operators  
\bea  \label{dc7}
\Delta C_7 (\mu_\smallw) & = & \f{\alpha_\smallw}{6\pi}\f{\as}{4\pi}
         \left [ s^2_\smallw \left(4 C_1 (x_t)   
        + \f{2}{3} H_0(x_t, x_z) 
  \right) 
        + \f{2}{3} s^4_\smallw \,\cZ_1 (x_t, x_z)  \right ] \, , \\  
\Delta C_8 (\mu_\smallw) & = & -\f{\alpha_\smallw}{3\pi}\f{\as}{4\pi}
 \Bigg [ s^2_\smallw \, 
H_0 (x_t, x_z) 
  +s^4_\smallw\, \cZ_1(x_t, x_z)\Bigg ] \, , 
\label{dc8}\\  
\Delta C_9 (\mu_\smallw) & = & -\f{\alpha_\smallw}{3\pi}\f{\as}{4\pi}
\Bigg [  B^{u}_1 (x_t) -  B^{d}_1 (x_t)   + 2 C_1 (x_t)   
        - \f{1}{3} G_0(x_t, x_z)  \label{dc9}   \\  
        & & \hspace{1.75cm} - s^2_\smallw \left(2 C_1 (x_t)   
        - \f{1}{3} G_0(x_t, x_z)  + \f{1}{3}\cZ_1(x_t, x_z) \right)+ 
      \f{1}{3}\, s^4_\smallw \cZ_1(x_t, x_z)\non   
        \Bigg ] \, , \\ 
\Delta C_{10}(\mu_\smallw) & = & - \f{\alpha_\smallw}{3\pi}\f{\as}{4\pi}
 \Bigg [  \widetilde{B}^{u}_1 (x_t)   
        -  \widetilde{B}^{d}_1 (x_t)   
        +  G_0 (x_t, x_z)  - s^2_\smallw\Big ( G_0 (x_t, x_z)   
        -  \cZ_1 (x_t, x_z) \Big) \non \\  
        & & \hspace{1.75cm} - s^4_\smallw    \, \cZ_1 (x_t, x_z) \label{dc10}
        \Bigg ] \, . 
\eea 
As we have seen above, there are  also  contributions to $C_{3,4}$
which can be extracted from (\ref{Heffz2}), (\ref{finZ}),
(\ref{finBu}) and (\ref{finBd}). However, any \ew correction to a 
gluon penguin diagram would contribute at the same order. The subset 
of diagrams we have computed is insufficient for these coefficients. 
We have organized the results in (\ref{dc7})-(\ref{dc10}) according to
powers of $s_\smallw$.\footnote{Of course, the argument $x_z = (1 -
s_\smallw^2)^{-1}$ of the functions $\cZ_0$ and $\cZ_1$ should also be
expanded in powers of $s_\smallw$. However, in our approximation the
whole $\ord(\mt^2)$ term of $\cZ_0$ of (\ref{Zhte}) has to be included
and can therefore be absorbed in the first term of the $s_\smallw$
expansion. Once this is done, expanding $x_z$ becomes numerically
irrelevant.} It should be clear by now that the zeroth order
coefficient is complete and gauge-invariant. The same applies to the 
coefficient of $s_\smallw^4$, as the only missing part of our 
calculation --- the QCD corrections to the photon penguin diagrams ---
is of  $\ord(\alpha_\smallw s^2_\smallw)$ and cannot contribute to
it. On the other hand, only the leading term of the HTE of the 
$s^2_\smallw$ coefficient is complete and gauge-invariant. 

\begin{comment}{\Delta C_3 (\mu_\smallw) & = &  -\f{\alpha_\smallw}{6 \pi}
 \f{\as}{4\pi} \Bigg [    
         B^{u}_1 (x_t)         + 2 B^{ d}_1 (x_t)   
        - C_1 (x_t)   
        + \f1{6} G_0(x_t, x_z)  +\f{s^2_\smallw}{6}\cZ_1(x_t, x_z) \Bigg ] \\  
\Delta C_4 (\mu_\smallw)&=& -\f{\alpha_\smallw}{6\pi}\f{\as}{4\pi}\Bigg [ 
         \widetilde{B}^{u}_1 (x_t)   
        + 2 \widetilde{B}^{d}_1 (x_t)   
        - \f{1}{2} G_0(x_t, x_z)  -\f{s^2_\smallw}{2}\cZ_1(x_t,x_z)\Bigg ] \\  
}\end{comment}
As a first check of our results, we can verify 
 that the dependence of the NLO coefficients $C_i$ on the matching scale $\mu_\smallw$ and 
on the top mass renormalization scale $\mu_t$ is removed 
by the scale dependence of the calculated NNLO corrections, up to terms
$\ord(\alpha_\smallw s^2_\smallw)$ originated by the missing photon-penguins.
 Indeed, it is straightforward to see that
\be
x_{\mu_t}\f{\partial}{\partial x_{\mu_t}}  \left(
\vec C_e^{(1)}(\mw)  + \f{\as}{4\pi}\vec 
 C_{es}^{(2)}(\mw) \right) =\ord(\as^2)
\ee
and similarly for the $\mu_\smallw$ dependence.
This follows from 
\be x_{\mu_t}\f{\partial}{\partial x_{\mu_t}}\vec C_e^{(1)}(\mw)=
-\gamma_0^m \,\f{\as}{4\pi}\, x_t \,\f{\partial}{\partial x_t}
\vec C_e^{(1)}(\mw) 
\ee
and 
\be
x_{\mu_\smallw}\f{\partial}{\partial x_{\mu_\smallw}} \vec C_e^{(1)} (\mu_\smallw)=
-\f1{2}\, \f{\as}{4\pi}\, \hat\gamma^{(0)^T} \,\vec  C_e^{(1)}(\mu_\smallw) .
\ee
 Here
$\gamma_0^m=8$ and $\hat\gamma^{(0)}$ are
the LO anomalous dimension of the top mass and the LO 
anomalous dimension matrix of the operators $Q_i$. Additional $\mu_\smallw$ 
dependent contributions of $\ord(\alpha_\smallw \as s^2_\smallw)$
come from the QED induced mixing between the gluon and \ew penguin
operators.

The numerical values of the Wilson coefficients at the \ew
scale are reported in Table 1, where we compare the NLO and NNLO results.
In all numerical calculations we employ $\mw=80.394$ GeV, $\mz=91.1867 $ GeV, 
and $\as(\mw)=0.121$ \cite{LEPEWWG}.
 In Table 1 we furthermore fix $\mu_\smallw=\mu_t=\mw$, 
and consequently adopt $\mt(\mw)=175.5$ GeV, which follows from the
experimental value of the pole top mass, $\mt=174.3\pm 5.1$ GeV. For
the \ew mixing angle we use $s^2_\smallw=\sin^2\hat{\theta}_{\msbar}(\mz)\simeq 0.23145$
 \cite{LEPEWWG,dgs}.
\renewcommand{\arraystretch}{1.15}
\begin{table}[t] 
\[
\begin{array}{|c|r|r|r|r|r|r|}\hline
C_i(\mw)  & {\rm NLO}_{\rm NDR} & {\rm NNLO}^{(1)}_{\rm NDR}   & {\rm
NNLO}^{(2)}_{\rm NDR} & {\rm NNLO}^{\rm HTE}_{\rm NDR} &{\rm NLO}_{\rm HV}
& {\rm NNLO}^{(2)}_{\rm HV} \\  \hline\hline
C_7(\mw) & 0.135     & 0.115      &  0.116 & 0.114 & 0.158 & 0.142\\ \hline
C_8(\mw)  & 0        &  0.001 
&  0.002 
& 0.002 
& 0 & -0.005
\\ \hline
C_9(\mw)  &-1.091     & -1.004    &  -1.002 & -1.014 & -1.067& -0.963 \\ \hline
C_{10}(\mw)  & 0        &  -0.019   &  -0.024 & -0.019 & 0& -0.003\\ \hline
\end{array}            
\]
\caption{\sf 
Wilson coefficients of the \ew penguin operators at the scale $\mw$ in units
$\alpha$ in  the NDR scheme and HV schemes (see text).
}\label{ta1}
\end{table}
We give three different values for the NNLO coefficients in the NDR scheme:
${\rm NNLO}^{(1)}$ corresponds to the expressions given in
(\ref{dc7})-(\ref{dc10}), which, as mentioned above, contain some
gauge-dependent terms calculated in the $\xi_\smallw=1$ gauge. 
In ${\rm NNLO}^{(2)}$,  instead, we expand the $s^2_\smallw$
coefficients of (\ref{dc7})-(\ref{dc10}) in inverse powers of $\mt$ and retain
only the leading HTE component. To this end we recall that
\be
C_1^{\rm HTE}(x_t)= x_t \left(\f{4}{3} -\zeta(2)+ \ln x_{\mu_t} - \ln x_t \right).
\ee
The formulation ${\rm NNLO}^{(2)}$ is strictly gauge-independent.
The QCD corrections modify $C_{7,9}(\mw)$ by about $-15\%$ and $+8\%$,
respectively. The difference between ${\rm NNLO}^{(1)}$ and ${\rm NNLO}^{(2)}$ is very small,
which is consistent with our expectations  about the contributions 
of the QCD corrected photon penguin diagrams. 
The inclusion of a recent
result  for the color-singlet photon penguin contribution \cite{misiak}
would change the results for $C_7(\mw)$ very little (by about $+3\%$)
and  marginally ($-0.4\%$) for  $C_{9}(\mw)$. Following
 Section\,\ref{strategy}, we will not include it here.
Finally, the fourth column of Table 1 gives
$C_{7-10}(\mw)$ in the NDR scheme for  the case in which
all $\ord(\as)$ corrections are calculated at
leading order in the HTE, ${\rm NNLO}^{\rm HTE}$. The agreement with the third
 column is relatively good also in this case.
We also observe that in the $\xi_\smallw = 1$ gauge and at $\mu_t=\mw$ the dominant NNLO 
contribution to $\Delta C_{7-10}$ is provided by $C_1(x_t)$, the color-singlet
corrections to the $Z^0$-penguin diagrams.

The QCD corrections of (\ref{dc7})-(\ref{dc10}) are specific to the 
NDR scheme.  Using the results given in Section\,\ref{Zpeng},
it is not difficult to find the expressions for the Wilson 
coefficients of (\ref{dc7})-(\ref{dc10}) in the HV scheme: 
the cofactors of $C_0(x_t)$
in (\ref{G0}), (\ref{H0}) become (1,5) instead of 
(5,-7) in NDR; the cofactors of $B_0$ in $B^u_1$ and $\widetilde{B}^u_1$
in (\ref{Bu1}), (\ref{Bu1t}) are (6,$-2$) instead of ($-38/3,-10$); 
the cofactors of $B_0$ in $B^d_1$ and $\widetilde{B}^d_1$
in (\ref{Bd1}), (\ref{Bd1t}) become $(-3/2,1/2)$  instead of ($19/6$,5/2).
The numerical values of $C_{7-10}(\mw)$ in the HV scheme at NLO and NNLO
for $\mu_t=\mu_\smallw=\mw$ 
are given in the last two columns of Table~1. Also in this scheme the
 QCD corrections to $C_{7, 9} (\mw)$ are $\ord(10\%)$. We note finally
that at NNLO $C_{8, 10} (\mw)$ become non-zero, but still are very small.

\subsection{Reduction of the $\mu_t$-dependence}
\begin{figure}[t]
\centerline{
\mbox{
\psfig{file=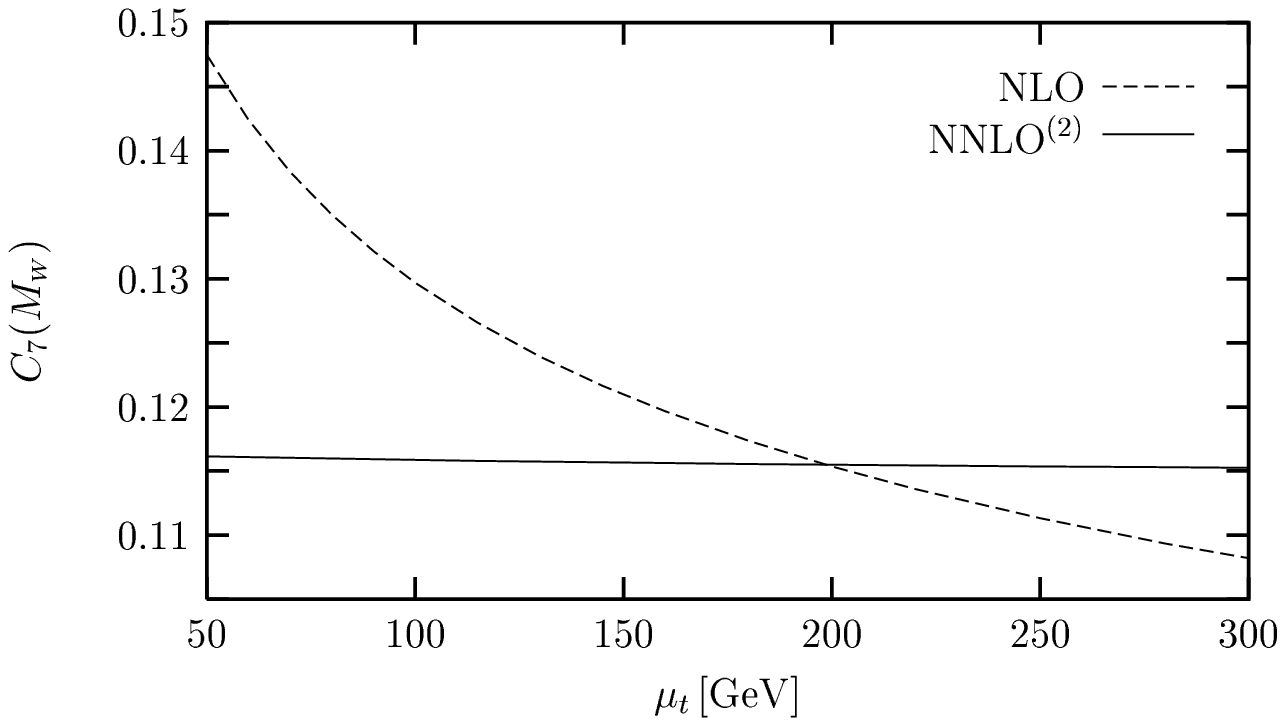,rheight=5.2cm,height=13cm,width=8.5cm
}}
\hspace{3mm}\mbox{
\psfig{file=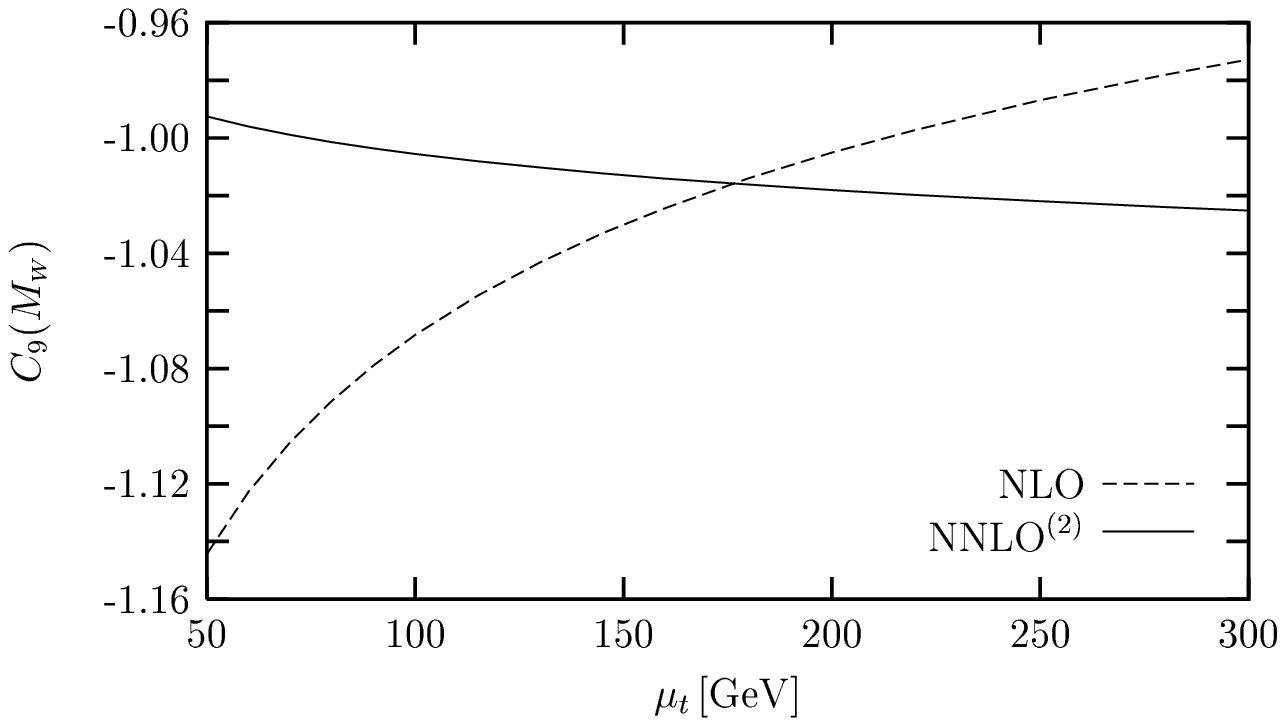 ,rheight=5.2cm,height=13cm,width=8.5cm
}}}
\caption{$\mu_t$ dependence of $C_7(\mw)$ and $C_9(\mw)$ at NLO 
and after the inclusion of
the NNLO corrections for $\mu_\smallw=\mw$  in the NDR scheme.} 
\label{scaledep}
\end{figure}

It is interesting to compare the $\mu_t$ dependence of the Wilson coefficients
before and after the inclusion of the $\ord(\as)$ corrections. This is done in
Figs. \ref{scaledep} for $C_7 (\mw)$ and $C_9 (\mw)$, where we have used the leading log expression for the 
running mass of the top
\be
\mt(\mu_t)= \mt(\mt) \left[ \f{\as(\mu_t)}{\as(\mt)}\right]^{12\over23}
\ee
and employed the ${\rm NNLO}^{(2)}$ expressions in the NDR scheme. 

Despite the fact that the NNLO corrections have not been computed completely,
the reduction of the scale dependence is remarkable, and is again consistent
with the idea that the contributions we have calculated are the dominant
ones. We also observe that the QCD corrections to $C_7 (\mw)$ and $C_9
(\mw)$ are particularly small for
$\mu_t\approx\mt$. This has also been found in the case of rare
semi-leptonic decays \cite{buchalla:93b}. As we will discuss below
this pattern does not apply to $C_8$ and $C_{10}$. However, in view of
the fact that $C_8 (\mw) = C_{10} (\mw) = 0$ at NLO, we will study
their $\mu_t$-dependence for $\mu \ll \mw$.  

In practical applications it is often useful to have simple and 
compact formulas
for the Wilson coefficients at the weak scale. For $\mu_\smallw=\mw$ and
$\mu_t=\mt$,  the NDR coefficients in the NNLO$^{(2)}$ formulation 
and in units of $\alpha$
can be written as
\bea
& & C_7(\mw)=0.02185 \, x_t^{1.1482} \, , \hspace{1cm}
C_8(\mw)=0.000718 \, x_t^{0.661} \, ,\non\\
& & C_9(\mw)=-0.438 \,x_t^{0.580}\, , \hspace{0.7cm}
C_{10}(\mw)=-0.004224 \,x_t^{1.1071} \, , 
\eea
which have to be compared with the NLO expressions
\be
C_7^{\rm NLO}(\mw)=0.02268\, x_t^{1.1423} \, , \ \ \ 
C_9^{\rm NLO}(\mw)=-0.434\, x_t^{0.590} \, , \ \ \ 
C_{8,10}^{\rm NLO}(\mw)=0 \, . 
\ee
These expressions reproduce the results of the complete formulas with an
accuracy of 0.2\% or better within two sigmas of the present $\mt$ value.

\subsection{RGE Evolution and Scheme Dependence}
\renewcommand{\arraystretch}{1.15}
\begin{table}[t] 
\[
\begin{array}{|c|r|r|r|r|}\hline
C_i  & { \rm NLO_{NDR}} & {\rm NNLO_{NDR}}   & { \rm NLO_{HV}} &{ \rm NNLO_{HV}}
 \\  \hline\hline
C_7(\mu_b) & -0.002     & -0.011      &  -0.002 & -0.010 \\ \hline
C_8(\mu_b)  & 0.055        &  0.060     &  0.061  & 0.050 \\ \hline
C_9(\mu_b)  & -1.336     &  -1.218  &  - 1.336 & -1.243  \\ \hline
C_{10}(\mu_b)  & 0.277        &  0.209  &  0.280 & 0.260 \\
\hline\hline
C_7(\mu_K) &-0.030& -0.032 & -0.028& -0.037 \\ \hline
C_8(\mu_K)  & 0.142 & 0.160 & 0.151& 0.135\\ \hline
C_9(\mu_K) &-1.538& -1.375 & -1.538& -1.445\\ \hline
C_{10}(\mu_K)& 0.582& 0.441&0.589 & 0.553\\ \hline
\end{array}            
\]
\caption{\sf 
Wilson coefficients of the \ew penguin operators 
at typical hadronic scales
$\mu_b=4.4$ GeV and $\mu_K=1.3$ GeV in units of
$\alpha$ in  the NDR scheme and HV schemes for $\mu_\smallw=\mu_t=\mw$ (see text).
}\label{ta2}
\end{table}
Let us now study the evolution of the coefficients down to a typical hadronic
scale. The inclusion of NNLO contributions proceeds as explained in
Section 3. We will consider two cases: the one of $B$ meson decays, for which
we will use $\mu_b=4.4$ GeV and the one of $K$ meson decay, corresponding to
$\mu_K=1.3$ GeV, as used in the analysis of $\varepsilon'/\varepsilon$ 
\cite{bosch99}. The results for NDR and HV schemes 
are shown in Table 2 for $\mu_t=\mu_\smallw=\mw$. 
We recall that  part (but not all) of the scheme 
dependence of $C_{7-10}$ is canceled in the evolution 
against similar terms in the anomalous dimension matrix (see e.g.\ 
\cite{buchalla:96}).
As demonstrated in detail in Section 4 the renormalization scheme
dependence of $C_{7-10} (\mw)$ discussed above is canceled by the
first term in (\ref{eq:ii}) stemming from the renormalization group
transformation. The coefficients $\vec C (\mu)$ are however scheme
dependent through the scheme dependence at the lower end of the RGE
evolution represented by the second term in (\ref{eq:ii}).
The complete cancelation of the scheme dependence of physical
amplitudes occurs only with the inclusion  of the matrix elements of the operators
$Q_i$. We employ $\as(\mz)=0.119$, corresponding 
approximately to $\Lambda^{(4)}=340$ MeV.
In Table \ref{ta2} the entries labeled by NLO refer to the NLO case described
in Section 2. The entries
identified by  NNLO, instead, correspond to our approximation of the full
NNLO result in the form  NNLO$^{(2)}$. 
At $\mu=4.4$ GeV the shifts due to the new contributions in the NDR scheme
and for $\mu_t=\mw$
are about $+9\%$ for $C_8$, $+9\%$ for $C_9$, $-24\%$ for $C_{10}$. 
$C_7$ remains very small.
At $\mu_K=1.3$ GeV the situation is similar, although the shifts are
naturally  more pronounced. In the case of the HV scheme the NNLO corrections
to $C_9$ are somewhat smaller than in the NDR scheme. They are comparable for
$C_7(\mu_b)$ and somewhat larger for $C_7(\mu_K)$. The strongest scheme
dependence is observed in the case of $C_8$ and $C_{10}$, which is not
surprising as $Q_8$ and $Q_{10}$ are color non-singlet operators. Whereas 
$C_8$ is enhanced in the NDR scheme, it is suppressed in the HV scheme.
$C_{10}$ is suppressed in both schemes but the effect is substantial 
in the NDR scheme and rather small in the HV scheme.

As explained in Section \,\ref{strategy}, 
there are other NNLO contributions that we have neglected.
Some of them are not known, but we can check the magnitude of 
the neglected $\ord(\alpha_\smallw \as s^2_\smallw)$ effects from the term 
$\as (\mw)/(4 \pi) \hat R^{(1)}(\mu,\mw)
\vec C_s^{(1)}$ in (\ref{cII}). It turns out that these effects are much
smaller than the NNLO contributions we have considered and are completely negligible.
We also notice that, among the NNLO contributions in (\ref{cI}), the one
proportional to $\vec C^{(2)}_{es}$ is by far the dominant in the
calculation of $C_7$ and $C_9$.

\begin{figure}[t]
\centerline{
\mbox{
\psfig{file=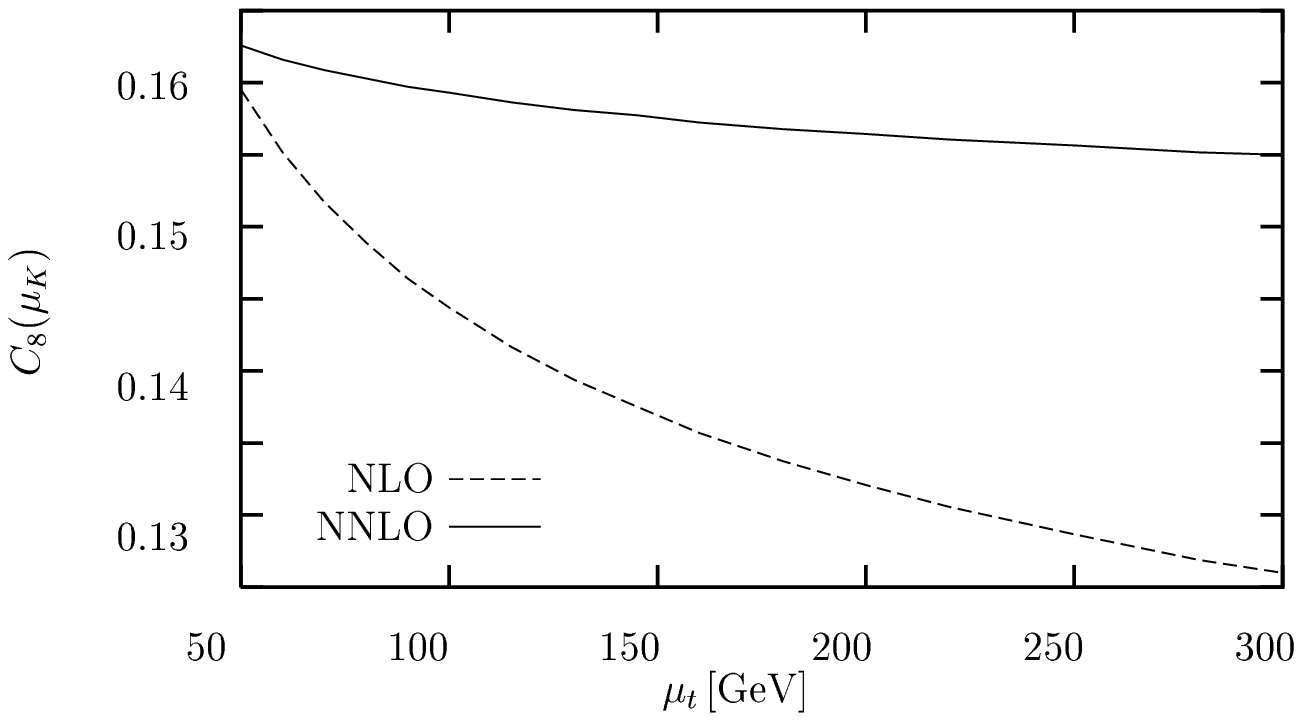,rheight=5.2cm,height=13cm,width=8.5cm}}
\hspace{3mm}\mbox{
\psfig{file=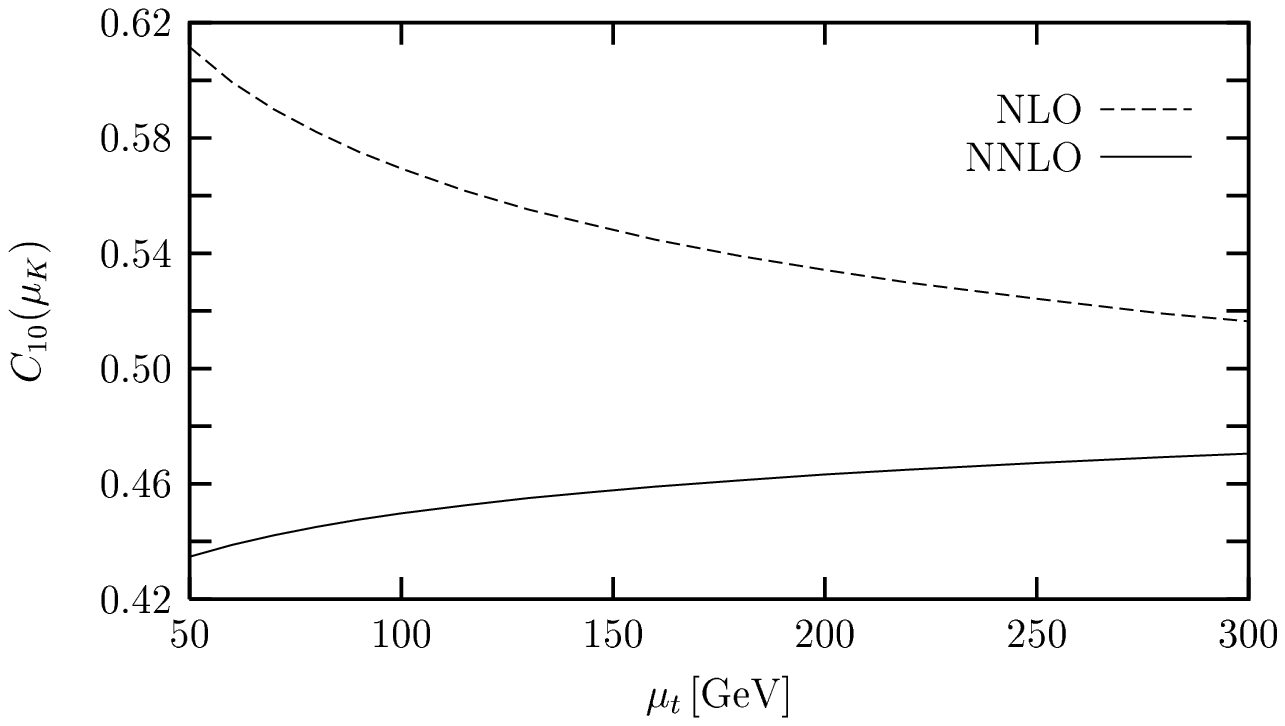 ,rheight=5.2cm,height=13cm,width=8.5cm}}}
\caption{$\mu_t$ dependence of $C_8(\mu_K)$ and $C_{10}(\mu_K)$  at NLO 
and after the inclusion of
the NNLO corrections for $\mu_\smallw=\mw$  in the NDR scheme. } 
\label{scaledepc8}
\end{figure}

Fig.\,\ref{scaledepc8} shows the $\mu_t$ dependence of $C_8(\mu_K)$ 
and $C_{10}(\mu_K)$ at NLO and NNLO order in the NDR scheme. 
Again, the reduction of the dependence on the
renormalization scale of the top mass is remarkable. In contrast to
$C_7$ and $C_9$ the NNLO corrections to $C_8$ and $C_{10}$ are
substantial in a large range of $\mu_t$ and a ``naive'' choice $\mu_t
= \mt$ in the NLO expressions would, in particular in the case of
$C_8$, totally misrepresent the true value of these coefficients. This
peculiar behaviour of $C_8$ and $C_{10}$ can be traced back to the
fact that $C_8 (\mw) = C_{10} (\mw) = 0$ at NLO. 

In Table \ref{ta3}  we show the results for the Wilson coefficients  as in
Table \ref{ta2} but this time choosing $\mu_t=\mt$. We observe a significant
reduction of the NNLO corrections in the case of $C_7$ and $C_9$ relative to
Table \ref{ta2}. The corrections to $C_8 $ and $C_{10}$ in the NDR scheme
increase and decrease, respectively. In the case of HV they are smaller 
than in Table \ref{ta2} but this time $C_{10}$ is slightly enhanced. In any
case the strong scheme dependence of $C_8 $ and $C_{10}$ observed in 
Table \ref{ta2} is also evident here. 

\begin{table}[t] 
\[
\begin{array}{|c|r|r|r|r|}\hline
C_i  & { \rm NLO_{NDR}} & {\rm NNLO_{NDR}}   & { \rm NLO_{HV}} &{ \rm NNLO_{HV}}
 \\  \hline\hline
C_7(\mu_b) & -0.009     & -0.011      &  -0.009 & -0.010 \\ \hline
C_8(\mu_b)  & 0.053        &  0.059     &  0.059  & 0.051 \\ \hline
C_9(\mu_b)  & -1.249     &  -1.241   &  -1.249 & -1.264  \\ \hline
C_{10}(\mu_b)  & 0.256        &  0.218  &  0.259 & 0.266 \\
\hline\hline
C_7(\mu_K) &-0.036& -0.033 & -0.034& -0.037 \\ \hline
C_8(\mu_K)  & 0.135 & 0.157 & 0.145& 0.135\\ \hline
C_9(\mu_K) &-1.437& -1.403 & -1.437& -1.468 \\ \hline
C_{10}(\mu_K)& 0.539& 0.459& 0.546 & 0.565\\ \hline
\end{array}            
\]
\caption{\sf 
Wilson coefficients of the \ew penguin operators at typical hadronic scales
$\mu_b=4.4$ GeV and $\mu_K=1.3$ GeV in units of
$\alpha$ in  the NDR scheme and HV schemes for 
$\mu_\smallw = \mw$, $\mu_t = \mt$ (see text). 
}\label{ta3}
\end{table}

\subsection{Scheme Dependence of $C_8$ and $\varepsilon'/\varepsilon$}
The strong scheme dependence of $C_8$ at the NNLO level is welcome. In the case
of the CP-violating ratio $\varepsilon'/\varepsilon$, the operator $Q_8$ is by 
far the most important electroweak penguin operator due to its large $\Delta I 
= 3/2$ matrix element $\langle Q_8 (\mu_K) \rangle_2 = B_8^{(3/2)} (\mu_K) 
\langle Q_8 (\mu_K) \rangle_2^{\rm{vac}}$, where "$\rm{vac}$" stands for the vacuum 
insertion approximation. The scheme dependence of $\langle Q_8 (\mu_K) \rangle_2$
resides fully in $B_8^{(3/2)} (\mu_K)$. As the contribution of $Q_8$ is the 
dominant $\ord (\aem)$ contribution to $\varepsilon'/\varepsilon$ one expects 
that the product $B_8^{(3/2)} (\mu_K) C_8 (\mu_K)$ is approximately $\mu_K$ and 
renormalization scheme independent with small $\mu_K$ and scheme dependences 
to be canceled by contributions of other operators which mix with $Q_8$ under 
renormalization. This is supported by renormalization group studies
\cite{BJL93} which also 
show that at the NLO level $B_8^{(3/2)} (\mu_K) C_8 (\mu_K)$ is only weakly 
dependent on $\mu_K$ for $1 \gev \leq \mu_K \leq 2 \gev$.

The situation with the scheme dependence of $B_8^{(3/2)} (\mu_K) C_8 (\mu_K)$
is different. Only by including the NNLO corrections to $C_8 (\mu_K)$ 
calculated in the present paper $B_8^{(3/2)} (\mu_K) C_8 (\mu_K)$ turns out to 
be almost scheme independent, whereas a substantial scheme dependence is 
observed at NLO. Indeed using the results of Table 3 and $B_{8, \, \rm{HV}}^
{(3/2)} \simeq  1.2 \, B_{8, \, \rm{NDR}}^{(3/2)}$ \cite{bosch99} we find
for $\mu_K = 1.3 \gev$ 
\be \label{comp}
\frac{B_{8, \, \rm{HV}}^{(3/2)} (\mu_K) C_8^{\rm{HV}} (\mu_K)}{B_{8, \, 
\rm{NDR}}^{(3/2)} (\mu_K) C_8^{\rm{NDR}} (\mu_K)} = \left \{
\begin{array}{r@{\hspace{1cm}}l}
   1.29 & \mbox{NLO} \, , \\
   1.03 & \mbox{NNLO} \, .
\end{array}
\right.
\ee
This result can be understood by recalling that at NLO $C_8$ has the formal 
expansion $\ord (\aem/\as) + \ord (\aem)$. Now the NLO term $\ord (\aem)$ is
substantially larger than the leading term $\ord (\aem/\as)$ mainly due to the 
$Z^0$-penguin diagrams which contribute first at the NLO level. In evaluating 
numerically the product $B_8^{(3/2)} C_8$ one effectively includes a term 
$\ord (\aem \as)$ which originates in the product of the large $\ord (\aem)$ 
NLO term in $C_8$ and the scheme dependent $\ord (\as)$ correction in 
$B_8^{(3/2)}$. As the $\ord (\aem \as)$ term in question is really a part of 
the NNLO contribution and moreover it is substantial, the resulting scheme 
dependence of $B_8^{(3/2)} C_8$ at NLO is large. Including the 
$\ord (\aem \as)$
corrections to $C_8$ removes this scheme dependence to a large extent as seen
in (\ref{comp}).    

We would like to remark that the corresponding product $B_6^{(1/2)} C_6$ 
related to the dominant QCD-penguin operator in $\varepsilon'/\varepsilon$, 
exhibits a much smaller scheme dependence at NLO than $B_8^{(3/2)} C_8$. In 
this case the ratio corresponding to (\ref{comp}) is found to be $1.08$ at 
NLO. This is related dominantly to the fact that the NLO contribution to 
$C_6$ is relatively small compared to the leading term in contrast to the case 
of $C_8$ as discussed above.

What is the impact of our results for $C_8$ on
$\varepsilon'/\varepsilon \,$? 
Clearly the main theoretical uncertainties in $\varepsilon'/\varepsilon$
reside in the values of the hadronic matrix elements which are substantially 
larger than the renormalization scheme uncertainties just discussed. Yet
our calculation of NNLO corrections allows us 
to reduce considerably the $\mu_t$ and in particular the 
renormalization scheme dependence in the electroweak penguin
sector. However, in order to give a shift in $\varepsilon'/\varepsilon$ due to NNLO 
corrections one would have to include similar corrections to QCD-penguin 
contributions and subdominant $\ord (\aem \as)$ terms.

On the other hand, the inspection of Table 3 and (\ref{comp}) shows that the 
role of the electroweak penguins for fixed hadronic matrix elements is
increased by roughly $16 \%$ in the NDR scheme and decreased by
roughly $7 \%$ in the HV scheme compared to the NLO results. As
electroweak penguins contribute negatively to
$\varepsilon'/\varepsilon$, which is dominated by a positive
contribution from the QCD penguin operator $Q_6$, the NNLO corrections
to $C_8$ calculated here suppress
$(\varepsilon'/\varepsilon)_{\rm{NDR}}$ and enhance
$(\varepsilon'/\varepsilon)_{\rm{HV}}$ over their NLO values. As an
example taking central values of the parameters used in \cite{bosch99}
and including NNLO corrections to $C_{7-10} (\mu_K)$ we find 
\be
\varepsilon'/\varepsilon = \left \{
\begin{array}{r@{\hspace{1cm}}l}
   5.9 \cdot 10^{-4} & \mbox{NDR} \, , \\
   6.3 \cdot 10^{-4} & \mbox{HV} \, .
\end{array}
\right.
\ee
to be compared with $7.1 \cdot 10^{-4}$ (NDR) and $6.1 \cdot 10^{-4}$
(HV) at NLO. Here in contrast to \cite{bosch99} we have used $B_{8, \,
\rm{HV}}^{(3/2)} \simeq  1.2 \, B_{8, \, \rm{NDR}}^{(3/2)}$ and $B_{6,
\, \rm{HV}}^{(1/2)} \simeq  1.2 \, B_{6, \, \rm{NDR}}^{(1/2)}$ which
results in higher HV values than obtained there. We have checked that
the remaining scheme dependence resides dominantly in the QCD-penguin 
contributions for which NNLO corrections are unknown. For larger
(smaller) $B_8^{(3/2)}$ at fixed $B_6^{(1/2)}$ the impact on 
$\varepsilon'/\varepsilon$ coming from NNLO corrections to electroweak
penguin contributions is larger (smaller).

\subsection{Universality of the Functions $X$ and $Y$}
As discussed in the Introduction, any decay amplitude can 
be written as a linear combination of $\mt$-dependent functions present in the
initial conditions $C_i(\mw)$.
In the absence of QCD corrections the gauge independent set relevant for
non-leptonic and semi-leptonic rare $K$ and $B$ decays is given by 
\cite{PBE0}
\be
X_0 = C_0 - 4 B_0 \, , \ \ \ \ 
Y_0 = C_0 - B_0 \, , \ \ \ \
Z_0 = C_0 + \frac14 D_0
\ee
and $E_0$, with $C_0, B_0, D_0$ and $E_0$ entering $C_i(\mw)$
in (\ref{eq:CMw3})--(\ref{eq:CMw9}). Here we will only discuss $X_0$
and $Y_0$. In the case of semi-leptonic FCNC processes the inclusion
of $\ord(\as)$ corrections to $Z^0$-penguin and box diagrams generalizes $X_0$ 
and $Y_0$ to
\be
X_\ell (x_t) = C_\ell(x_t) - 4 B_\ell(x_t,+1/2), \ \ \ \ 
Y_\ell (x_t) = C_\ell(x_t)-B_\ell(x_t, -1/2),
\ee
where $C_\ell(x_t)\equiv C(x_t)$ is given in (\ref{eqsecv:cfunction}) and
\be
B_\ell(x_t, \pm1/2)=B_0(x_t) +\frac{\as}{4\pi} 
B_1(x_t, \pm1/2)
\ee
with $B_1(x_t, \pm1/2)$ given in \cite{misiak99,buras:99a}.
Concentrating first on the operators $Q_3$ and $Q_9$ and $\ord(\alpha_\smallw)$
terms in (\ref{eq:CMw3}) and (\ref{eq:CMw9}), respectively, our calculation of
gluonic corrections to box and $Z^0$-penguin diagrams provides the
generalization of $X_0$ and $Y_0$ relevant for non-leptonic decays as follows
\bea
\Delta C_3(\mw)&=& \frac{\alpha_\smallw}{6\pi}\left[2Y_0-X_0\right]
\hspace{2mm} \to
\hspace{2mm}\frac{\alpha_\smallw}{6\pi}\left[2Y_q-X_q\right] \, , \\
\Delta C_9(\mw)&=& -\frac{\alpha_\smallw}{6\pi}\left[2Y_0+2X_0\right]
\hspace{2mm} \to \hspace{2mm}
-\frac{\alpha_\smallw}{6\pi}\left[2Y_q+2X_q\right] \, , 
\eea
where 
\bea
X_q (x_t) &=& X_0 (x_t)  +\frac{\as}{4\pi} \left( C_1(x_t)
-{1 \over 6} \, G(x_t, x_z) + B_1^u (x_t)\right) \, , 
\\
Y_q (x_t) &=& Y_0 (x_t) +\frac{\as}{4\pi} \left( C_1(x_t)
-{1 \over 6} \, G(x_t, x_z) - B_1^d (x_t)\right).
\eea
Analogously we can write in the case of the operators $Q_4$ and $Q_{10}$
\bea
\Delta C_4(\mw)=  \frac{\alpha_\smallw}{6\pi}\left[2\widetilde{Y}_q-
\widetilde{X}_q\right], \ \ \ \ \ \ \ \
\Delta C_{10}(\mw)=
-\frac{\alpha_\smallw}{6\pi}\left[2\widetilde{Y}_q+2\widetilde{X}_q\right]
\, , 
\eea
where 
\bea
\widetilde{X}_q (x_t) =\frac{\as}{4\pi} \left( {1 \over 2} \, G (x_t , x_z) +
\widetilde{B}_1^u (x_t) \right),\ \ \  \ \ 
\widetilde{Y}_q (x_t) = \frac{\as}{4\pi} \left( {1 \over 2} \, G (x_t,
x_z) - \widetilde{B}_1^d (x_t) \right).
\eea
Evidently, at NNLO in non-leptonic decays more $\mt$-dependent functions appear
than in the case of semi-leptonic FCNC processes. Moreover additional functions
are necessary to describe the $\mt$-dependence of the coefficients $C_7$ and
$C_8$ as seen in (\ref{dc7}) and (\ref{dc8}). Furthermore,
gluon corrections to photon penguins and \ew corrections to gluon penguins will
introduce new $\mt$-dependent functions not present in semi-leptonic
FCNC decays. 

We conclude therefore that at the NNLO level in non-leptonic decays  the
structure of $\mt$-dependence is much more involved than in semi-leptonic FCNC
decays. On the other hand, if we restrict our discussion to the dominant
$\mt$-dependence residing in $C_9(\mw)$ we can say something concrete about 
the violation of the universality of the $\mt$-dependent functions addressed
briefly in the Introduction. To this end we write
\bea
X_q (x_t) &\!\!\!\!=\!\!\!\!& X_\ell (x_t)  \!+\! \frac{\as}{4\pi} \! \left[
4 B_1(x_t,\!-1/2) \!-\! {1 \over 6} G (x_t, x_z) \!+\!
B_1^u (x_t) \right] \hspace{0mm} \equiv \hspace{0mm} \eta_{q\ell}^X\,
X_\ell (x_t) \hspace{0mm} \equiv \hspace{0mm} \eta_q^X \,X_0 (x_t), \hspace{1.2cm}
\\
Y_q (x_t) &\!\!\!\!=\!\!\!\!& Y_\ell (x_t)  \!+\! \frac{\as}{4\pi} \! \left[
B_1(x_t,\!-1/2) \!-\! {1 \over 6} G (x_t, x_z) \!-\!
B_1^d (x_t) \right] \hspace{0mm} \equiv \hspace{0mm} \eta_{q\ell}^Y\,
Y_\ell (x_t) \hspace{0mm} \equiv \hspace{0mm} \eta_q^Y \,Y_0 (x_t). 
\eea
Clearly, the size of the various $\eta$ factors depends on the choice of
$\mu_t$. In Table \ref{ta4} we report their  values for 
$\mu_t=\mw$ and $ \mu_t=\mt$ in the NDR and HV schemes. 
As usual, we fix $\mu_\smallw=\mw$. 
\renewcommand{\arraystretch}{1.25}
\begin{table}[t] 
\[
\begin{array}{|c|r|r|r|r|}\hline
 & \multicolumn{2}{|c|}{\mu_t=\mw} & \multicolumn{2}{|c|}{\mu_t=\mt}  \\ \hline
 & {\rm NDR} & {\rm HV}   & {\rm NDR} & {\rm HV} \\  \hline\hline
\eta^X_q & 0.912     & 0.894      &  0.980 & 0.962 \\ \hline
\eta^Y_q  & 0.911        &  0.908 & 1.006 & 1.003 \\ \hline
\eta^X_{q\ell}  & 0.985     & 0.968   &  0.986 & 0.966  \\ \hline
\eta^Y_{q\ell}  & 0.993        &  0.991   &  0.994 & 0.990 \\ \hline
\end{array}            
\]
\caption{\sf $\eta$ factors for the functions $X$ and $Y$ in different schemes
and for different $\mu_t$.
}\label{ta4}
\end{table}
For  $\mu_t=\mw$ the universality of $X$ and $Y$  is
broken at $\ord(\as)$ by  terms which are relatively small with respect to
the NNLO correction, although not negligible in the HV scheme.
This follows also from our
previous remark that for this choice of scale 
the largest contribution to $C_9$ 
comes from $C_1(x_t)$, which is the same for hadronic and semi-leptonic decays.
In the case of $\mu_t=\mt$, however,   the $\ord(\as)$ corrections 
to $X,Y$ never exceed $4 \%$ and $C_1(x_t)$ plays no longer 
 a dominant role. 
Although the universality of $X$ and $Y$ is broken
by effects which are of the same order of the NNLO correction, 
these corrections are anyway much smaller in this case for $X$ and $Y$
than in the case $\mu_t = \mw$.
\begin{comment}{
, we find $\eta^X_{q}=0.912$, $\eta^Y_{q}=0.911$,
$\eta^{X}_{q\ell}= 0.985$  and $\eta^{Y}_{q\ell}= 0.993$
 in the NDR scheme and 
$\eta^X_{q}=0.894$, $\eta^Y_{q}=0.908$,
$\eta^{X}_{q\ell}= 0.968$,  and $\eta^{Y}_{q\ell}= 0.991$  in the HV scheme.
To excellent approximation we have, 
using again $\mu_\smallw=\mw$, 
$\eta^X_{q}=0.980$, $\eta^Y_{q}=1.006$,
$\eta^{X}_{q\ell}= 0.986$  and $\eta^{Y}_{q\ell}= 0.994$
 in the NDR scheme and 
$\eta^X_{q}=0.962$, $\eta^Y_{q}=1.003$,
$\eta^{X}_{q\ell}= 0.966$,  and $\eta^{Y}_{q\ell}= 0.990$  in the HV scheme.
}\end{comment}

\section{Summary}
\label{concl}
In this paper we have calculated the $\ord(\as)$ corrections to the 
$Z^0$-penguin and \ew box diagrams relevant for non-leptonic 
$\Delta S=1$ and $\Delta B=1$ decays. This calculation provides the complete 
$\ord(\alpha_\smallw \as)$  and 
$\ord(\alpha_\smallw \as \sws \mt^2)$ corrections to the Wilson coefficients 
of the \ew penguin four quark operators relevant for non-leptonic 
$K$- and $B$-decays. We have given arguments supported by numerical estimates
that the corrections calculated by us constitute by far the dominant part of
the next-next-to-leading (NNLO) contributions to these coefficients in the
renormalization group improved perturbation theory.

The main results for $\ord(\as)$ corrections to the $Z^0$-penguin
diagrams can be found in  (\ref{Heffz2}) and
(\ref{finZ}). Those for the box diagrams in (\ref{finBu}) and
(\ref{finBd}). The main results for the Wilson coefficients of the
electroweak penguin operators are collected in
(\ref{dc7})--(\ref{dc10}). The numerical values of these coefficients
are collected in Tables 1--3 and in Figs. 5 and 6.

Our main findings are as follows: 
\begin{itemize}
\item[i)] The inclusion of NNLO corrections allows to reduce
considerably the uncertainty due to the choice of the scale $\mu_t$ in
the running top quark mass $\mt (\mu_t)$ present in NLO
calculations. This is illustrated in Figs. 5 and 6.
\item[ii)] While NNLO corrections to $C_7$ and $C_9$ are generally
moderate and very small for the choice $\mu_t = \mt$, they are sizable
in the case of $C_8$ and $C_{10}$. This is illustrated in Tables 5 and
6. In particular we observe substantial renormalization scheme
dependence in $C_8$ and $C_{10}$, whereas the scheme dependence in
$C_7$ and $C_9$ is significantly smaller. 
\item[iii)] The strong scheme dependence of $C_8$ allows to cancel to
a large extent the scheme dependence of the matrix element $\langle
Q_8 \rangle_2$ relevant for $\varepsilon'/\varepsilon$ so that the
contribution of this dominant electroweak operator to
$\varepsilon'/\varepsilon$ is nearly scheme independent. This should
be contrasted with the existing NLO calculations of
$\varepsilon'/\varepsilon$ which exhibit sizeable scheme dependence in
the electroweak penguin sector. 
\item[iv)] In the case of $\Delta B = 1$ decays the most important
among the electroweak penguin operators is the operator $Q_9$. As the
NNLO corrections for $\mu_t = \mt$ are in the ball park of a few
percent, our results have smaller impact on non-leptonic $\Delta B =
1$ decays except for the reduction of the $\mu_t$-dependence.
\item[v)] We have also investigated the breakdown of the universality
in the $\mt$-dependent functions $X$ and $Y$. As these functions are
dominated by the contribution of the color singlet $Z^0$-penguin
diagram which is universal, the breakdown of universality through
color non-singlet $Z^0$-contributions and box diagrams is small as
illustrated in Table 4. 
\end{itemize}

Although we have seen that there are arguments suggesting that
our subset of NNLO corrections is dominant, several other
contributions have to be calculated in order to complete the NNLO
analysis for non-leptonic decays. We have discussed
this formally in Section 3. A step in this direction has been made
recently in \cite{misiak} where $\ord (\as^2)$ corrections to the
initial values $C_{1-6} (\mw)$ have been calculated. Yet the complete 
$\ord (\as)$ corrections to the photon penguin diagrams relevant for
non-leptonic decays and in particular the three loop anomalous
dimensions $\ord (\as^3)$ and $\ord (\aem \as^2)$ of the set
$Q_{1-10}$ are unknown. The present work and the complementary
calculation in \cite{misiak} constitute the first steps torwards a
complete NNLO calculation of non-leptonic decays and we have
demonstrated here that the NNLO corrections to the Wilson
coefficients of electroweak penguin operators are of phenomenological
relevance. 

}\end{comment}

\subsection*{Acknowledgments}
We are grateful to E. Franco, S. J{\"a}ger, M. Misiak, 
L. Silvestrini and J. Urban for interesting discussions and
communications. A. J. B. would like to thank Fermilab theory group for
a great hospitality during his stay in September. This work has been
supported in part by the Bundesministerium f{\"u}r Bildung und Forschung
under contract 05 HT9WOA.

\begin{appendletterA}
\section*{Appendix}
In this Appendix we report the analytic expressions for 
the functions $\cZ_{0,1}(x_t, x_z)$ introduced in \eqs{W2loop}. 
They read 
\bea \label{Z0}
&&\hspace{-6mm}\cZ_0 (x_t, x_z)  =  -\f{x_t \left(20 - 20 x_t^2 - 457 x_z + 19 
        x_t x_z + 8 x_z^2 \right)}{32 (x_t - 1) x_z} \non \\
        & & + \f{x_t \left(10 x_t^3 - 11 x_t^2 x_z - x_t (30 - 16 x_z) 
        + 4 (5 - 17 x_z + x_z^2) \right)}{16 (x_t - 1)^2 x_z} \ln x_t 
        \non \\ 
        & & + \f{x_t \left(10 - 10 x_t^2 - 17 x_z - x_t x_z - 4 x_z^2 
        \right)}{16 (x_t - 1) x_z} \ln x_z - \f{x_z \left (10 x_t^2 - x_t 
          (4 - x_z) + 8 x_z \right)}{32 (x_t - 1)^2} \ln^2 x_t\non \\
        & &  - \f{1}{4} x_z^2 \ln^2 x_z
        - \left [\f{8 + 12 x_t + x_t^2}{4 x_z} - 
          \f{5 (x_t - 1)^2  (2 + x_t)}{16 x_z^2} \right. \\ 
        & &  \left. - \f{12 - 3 x_t^3 - 3 x_t^2 (4 - x_z) + 4 
        x_t (3 - x_z) + 4 x_z - x_z^2}{8 (x_t - 1)^2} \right ] \ln x_t 
        \ln x_z \non \\
        & & - \left(\f{8 + 12 x_t + x_t^2}{2 x_z} - \f{5 (x_t - 1)^2 
        (2 + x_t)}{8 x_z^2}
         - \f{3 \left (4 + 8 x_t + 2 x_t^2 - x_t^3 
        \right)}{4 (x_t - 1)^2} \right) \mbox{Li}_2 (1 - x_t)  
        \non \\
        & & + \f{(x_z - 1)^2 \left (5 - 6 x_z - 5 x_z^2 \right)}{4 
        x_z^2}\, \mbox{Li}_2 (1 - x_z) - \f{5 - 16 x_z + 12 x_z^2 + 2 
        x_z^4}{4 x_z^2} \,\zeta (2) \non \\ 
        & & + \f{x_t (4 - x_z) \left (88 - 30 x_z - 25 x_z^2 - 2 x_t 
        (44 - 5 x_z - 6 x_z^2) \right)}{32 (x_t - 1)^2 x_z} \,\phi 
        \left(\f{x_z}{4} \right) \non \\ 
        & & + \f{16 x_t^4 - x_t (20 - x_z) x_z^2 +8 x_z^3- 8 x_t^3 (14 +  
        5 x_z) + 8 x_t^2 (12 - 7 x_z + x_z^2)}{32 (x_t - 1)^2 x_z}\,  
        \phi \left (\f{x_z}{4 x_t} \right)  \non\\  
        & & - \Bigg [\f{22 + 33 x_t - x_t^2}{16 (x_t - 1) x_z}  
        - \f{5 (x_t - 1) (2 + x_t)}{16 x_z^2}  
        + \f{2 + 5 x_t^2 + 10 x_z + x_t (15 + 
        x_z)}{16 (x_t - 1)^2} \Bigg ] \Phi (x_t, x_z)\non       
\eea 
and 
\bea\label{Z1}
&& \hspace{-6mm}\cZ_1 (x_t, x_z)= 
         \f{x_t (20 - 20 x_t^2 - 265 x_z + 67 x_t 
        x_z + 8 x_z^2 )}{48 (x_t - 1) x_z} \non \\ 
        & & - \f{x_t \left(10 x_t^3 - 15 x_t^2 x_z + 4 (5 - 7 x_z + 2 
        x_z^2) - x_t (30 + 20 x_z + 4 x_z^2) \right)}{24 (x_t - 1)^2 x_z} 
        \ln x_t \non\\
          & & -  \f{x_t(10 - \! 10 x_t^2 - \!33 x_z + \!15 x_t x_z -\! 4 
        x_z^2 )}{24 (x_t - 1) x_z} \ln x_z
        + \f{x_z  (8 -\! 16 x_t + \!2 x_t^2 +\! 10 x_z + 7 x_t x_z 
       )}{48 (x_t - 1)^2} \ln^2 x_t \non\\ 
        & &+ \f{x_z (4 + 5 x_z)}{24} 
        \ln^2 x_z  + \Bigg [\f{20 + 6 x_t + x_t^2}{12 x_z}  
        - \f{5 (x_t - 1)^2 (2 + x_t)}{24 x_z^2}  \\
        & & \hspace{0.6cm} + \f{3 x_t^3 + 2 x_t^2 (12 - x_z) - x_t (18  
        - 16 x_z + x_z^2) - 2 (9 + 4 x_z - x_z^2)}{12 (x_t - 1)^2} 
        \Bigg ] \ln x_t \ln x_z \non\\ 
        & & + \left(\f{20 + 6 x_t + x_t^2}{6 x_z} - \f{5 (x_t - 1)^2 
        (2 + x_t)}{12 x_z^2} - \f{6 + 6 x_t - 8 x_t^2 - x_t^3}{2 (x_t - 
        1)^2} \right) \mbox{Li}_2 (1 - x_t) \non\\ 
        & & - \f{(x_z - 1)^2 \left (5 - 10 x_z - 7 x_z^2 \right)}{6 
        x_z^2} \,\mbox{Li}_2 (1 - x_z) 
         + \f{10 - 40 x_z + 36 x_z^2 + 4 x_z^3 + 5 x_z^4}{12 
        x_z^2} \,\zeta (2) \non\\ 
        & & + \f{x_t (x_z - 4) \left (24 - 26 x_z - 13 x_z^2 - 6 x_t  
        (4 - x_z - x_z^2) \right)}{16 (x_t - 1)^2 x_z} \,\phi \left 
        (\f{x_z}{4} \right)-\Bigg[\f{2 x_t^2 (2+x_t)}{3 (x_t - 1) x_z} \non\\ 
        & & \hspace{0.6cm} - \f{24 x_t^3 + 12 x_t^2 (14 + x_z) - 2 x_z  
        (4 + 5 x_z) - x_t (80 - 36 x_z + 7 x_z^2)}{48 (x_t - 1)^2} 
        \Bigg ] \,\phi \left(\f{x_z}{4 x_t} \right) \non\\ 
        & & + \Bigg [\f{10 - x_t - x_t^2}{8 (x_t - 1) x_z} - \f{5 (x_t - 1) 
        (2 + x_t)}{24 x_z^2}  
      + \f{6 +  3x_t^2 + 14 x_z + 5 x_t (7 + 
        x_z)}{24 (x_t - 1)^2} \Bigg ] \,  \Phi (x_t, 
        x_z) \, .  \non
\eea   
 We recall that $\zeta(2) = \pi^2/6$.
The function $\phi (z)$  appearing in the above expressions is given by
\bea
\phi(z) =\left\{
       \begin{array}{ll}
       4 \sqrt{{z \over 1-z}} ~\mbox{Cl}_2 ( 2 \arcsin \sqrt z ) \, , &  0 < z
\leq 1 \, , \\
  { 1 \over \beta} \left[ - 4 {\rm Li_2} ({1-\beta \over 2}) +
       2 \ln^2 ({1-\beta \over 2}) - \ln^2 (4z) +2\zeta(2) \right] \, ,  
        &z >1 \,,
       \end{array}\right.\nonumber
\eea 
where $\mbox{Cl}_2(x)= {\rm Im} \,{\rm Li_2} (e^{ix})$ is the Clausen function
and $\beta = \sqrt{1 - {1/ z}}$.  
Defining 
\be 
\lambda(x,y)  = \sqrt{(1 - x - y)^2 - 4 x y},\non
\ee
the function $\Phi(x,y)$ admits two different representations,
 according to the sign of 
$\lambda^2(x,y)$. For $\lambda^2\ge 0$ we have 
\bea
\label{phi2a} 
\Phi (x, y) =\lambda\left\{2\ln\left (\f{1 +x- y-\lambda}{2} \right ) 
        \ln\left ( \f{1 - x + y - \lambda}{2} \right ) - \ln x \ln y \right. 
\hspace{1cm} \nonumber \\  
        \left. -2 \, \mbox{Li}_2 \left ( \f{1 + x - y - \lambda}{2} \right ) - 
2\,\mbox{Li}_2\left(\f{1 -x+y -\lambda}{2} \right ) + 2 \,\zeta (2) \right \},
\eea
while for $\lambda^2\le 0$
\bea
\label{phi2b} 
\Phi (x, y) = -2\,\sqrt{-\lambda^2} \left \{ \mbox{Cl}_2 \left [2 \arccos 
     \left ( \f{-1 + x + y}{2 \sqrt{x y}} \right ) \right ] \right. \hspace{4
cm} \nonumber \\
   \left. + \mbox{Cl}_2 \left [2 \arccos \left ( \f{1 + x - y}{2 \sqrt{x}} 
\right ) \right ]
   + \mbox{Cl}_2 \left [2 \arccos \left ( \f{1 - x + y}{2 \sqrt{y}} \right ) 
\right ] \right \}.
\eea
Additional details on this function can be found in \cite{davydychev}.

\end{appendletterA}


\end{document}